%% file: main.tex
\documentclass[a4paper,11pt]{article}
\pdfoutput=1
\usepackage{jheppub}

\usepackage{amsmath}
\usepackage{mathtools}
\usepackage{multirow}
\usepackage{array}
\usepackage{bm}

\usepackage{graphicx}
\usepackage{lscape}
\usepackage{subfig}

\usepackage{hyperref}
\usepackage{booktabs}
\usepackage{float}

\usepackage{xspace}

\newcommand{\nn}{\nonumber \\}

\def\eps{\epsilon}
\def\top #1{\mathcal{T}_{#1}}

\allowdisplaybreaks[1]

\newcommand{\dA}{\ensuremath d\AA}
\newcommand{\Den}{\ensuremath D}
\newcommand{\dd}{\ensuremath \mathrm{d}}
\DeclareMathOperator{\dlog}{\mathit{d}log}

\newcommand{\FF}{\ensuremath \text{F}}
\newcommand{\FFvec}{\ensuremath \vec{\text{F}}}
\newcommand{\GG}{\ensuremath \text{I}}
\newcommand{\GGvec}{\ensuremath \vec{\text{I}}}

\newcommand{\MM}{\ensuremath \mathbb{M}}
\renewcommand{\AA}{\ensuremath \mathbb{A}}

\newcommand{\pavia}{INFN Sezione di Pavia, Via Agostino Bassi 6, 27100 Pavia, Italy}

\newcommand{\ub}{Department of Physics, University at Buffalo, The State University of New York, Buffalo 14260, USA}

\def\dlog{d\log}

\allowdisplaybreaks

\author[a]{Syed Mehedi Hasan,}
\author[b]{Ulrich Schubert}

\affiliation[a]{\pavia}
\affiliation[b]{\ub}

\emailAdd{syedmehe@pv.infn.it}
\emailAdd{ulrichsc@buffalo.edu}

\title{Master Integrals for the mixed QCD-QED corrections to the Drell-Yan production of a massive lepton pair}


\abstract{
We showcase the calculation of the master integrals needed for the two loop mixed QCD-QED virtual corrections to the neutral current Drell-Yan process $(q\bar{q}\rightarrow l^+ l^-)$. After establishing a basis of 51 master integrals, we cast the latter into canonical form by using the Magnus algorithm. The dependence on the lepton mass is then expanded such that potentially large logarithmic contributions are kept. After determining all boundary constants, we give the coefficients of the Taylor series around four space-time dimensions in terms of generalized polylogarithms up to weight four.
}

\begin{document}

\maketitle
\flushbottom
\input{tex/introduction}
\input{tex/notation}
\input{tex/deq}
\input{tex/conclusions}

\section*{Acknowledgments}
The authors would like to thank Roberto Bonciani, Pierpaolo Mastrolia, Doreen Wackeroth and Ciaran Williams for useful discussion.  US is supported by the National Science Foundation awards PHY-1719690 and PHY-1652066. Support provided by the Center for Computational Research at the University at Buffalo. SMH carried out part of the project at the University at Buffalo, supported in part by the National Science Foundation under grant no. NSF PHY-1719690.

\bibliographystyle{JHEP}
\bibliography{references}

\end{document}

%% file: tex/introduction.tex
\section{Introduction}

\begin{figure}
\begin{center}
\subfloat[Neutral Current]{\includegraphics[width = 3in]{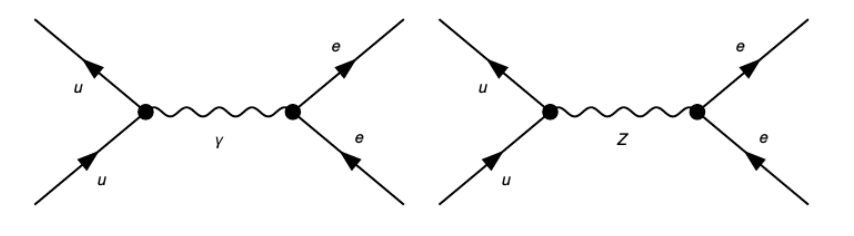}} 
\subfloat[Charged Current]{\includegraphics[width = 2in]{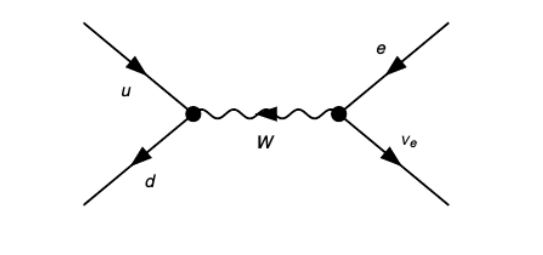}}
\caption{Feynman diagrams for the partonic Neutral Current and Charged Current DY processes at LO. The diagrams have been generated using Feynarts~\cite{Hahn:2000kx}.}
\label{fig:DYtree}
\end{center}
\end{figure}

The Drell-Yan (DY) processes, depicted at leading order in figure~\ref{fig:DYtree}, have big cross sections and clean experimental signatures, and thus are one of the best-studied processes at the LHC. In particular they can be used to determine important parameters in the electroweak (EW) sector like the weak mixing angle and W boson mass~\cite{Aaboud:2017svj,Sirunyan:2018swq}. The DY processes are also playing an important role as standard candles for the LHC in the form of luminosity measurements and detector calibrations. Furthermore the abundance of clean data make the DY processes a perfect place to determine parton distribution functions and search for Beyond Standard Model (BSM) physics. All these applications rely on a precise theoretical description of the DY processes, making it crucial for the physics program at the LHC.

The perturbative corrections to the Drell-Yan processes can be divided into two classes. The pure QCD corrections only occur in the initial state of the DY processes, due to the colorless nature of the leptonic final state. These corrections are known differentially up to next-to-next-to leading order (NNLO) ~\cite{Matsuura:1988sm,Hamberg:1990np} and inclusively at next-to-next-to-next-to-leading order (NNNLO)~\cite{Duhr:2020seh}.
In contrast the EW corrections can involve both the quarkonic initial and leptonic final state. These corrections have been computed at next-to-leading order (NLO)~\cite{Altarelli:1979ub,Wackeroth:1996hz,Baur:1998kt,Dittmaier:2001ay,Baur:2004ig,Zykunov:2006yb,Arbuzov:2005dd,CarloniCalame:2006zq,Baur:1997wa,Zykunov:2005tc,CarloniCalame:2007cd,Arbuzov:2007db} 
and there is an ongoing effort to extend the computation to NNLO~\cite{Degrassi:2003rw,Actis:2006ra,Actis:2006rb,Actis:2006rc}.

Starting at NNLO the EW and QCD corrections start to mix and are currently assumed to be the largest unknown correction in the high energy region~\cite{Campbell:2016dks}. These mixed corrections can be further divided according to the number of vector boson exchanges. While the factorizable contributions are characterized by a single vector boson exchange between the initial and final state, the non-factorizable contributions involve the exchange of two or more bosons. Among the factorizable Feynman diagrams, the mixed double virtual corrections were computed for the W and Z boson decay~\cite{Czarnecki:1996ei,Kara:2013dua} and the Z boson production~\cite{Kotikov:2007vr}. The double real contribution to the total cross section for the on-shell single gauge boson production has been presented in~\cite{Bonciani:2016wya} and the $\mathcal{O}(\alpha\alpha_s)$ corrections to the total partonic cross section of the process $q\bar{q}\rightarrow Z + X$ is calculated in~\cite{Bonciani:2019nuy}.

For the non-factorizable contributions, significant work has been done in the QCD-QED sector~\cite{Kilgore:2011pa,deFlorian:2018wcj,deFlorian:2015ujt,Blumlein:2020jrf,Ablinger:2020qvo} and by adopting the pole approximation~\cite{Dittmaier:2014qza,Dittmaier:2014koa,Dittmaier:2015rxo,Dittmaier:2016egk}. 
Nevertheless BSM physics might also show up in regions outside the resonance and therefore it is important to have control over the non-factorizable corrections beyond the pole approximation. Currently all ingredients, including the master integrals~\cite{Bonciani:2016ypc,vonManteuffel:2017myy,Heller:2019gkq}, for the construction of the amplitudes in the zero lepton mass limit are known.

However there are observable effects due to the lepton masses, when considering collinear radiation of photons. While the KLN theorem ensures that for a fully inclusive observable, all these effects cancel out, the use of lepton identification cuts breaks the fully inclusive nature of the measured cross sections~\cite{Baur:1997wa}. These analysis cuts give rise to large logarithmic contributions of the form $\log(s/m_l^2)$, which, due to their photonic origins, arise only from the QED part of the EW corrections~\cite{Baur:2001ze,Dittmaier:2009cr}. Therefore we can fully capture these logarithmic contributions by only considering the QED part of this amplitude for non-zero lepton masses. The calculation of the latter requires master integrals with a non zero lepton mass, which are the main subject of this publication.

Fortunately the huge number of multi scale integrals appearing in loop amplitudes are linearly dependent through integration-by-parts identities (IBPs)~\cite{Tkachov:1981wb,Chetyrkin:1981qh,Laporta:2001dd}. Therefore this huge number of integrals can be expressed in terms of a much smaller set of basis integrals called master integrals. Interestingly the IBPs can be also employed for the solutions of these master integrals, since their derivatives in respect to the kinematic invariants lay within the same space that is spanned by the IBPs. The resulting first order differential equations~\cite{Kotikov:1990kg,Remiddi:1997ny,Gehrmann:1999as} can then be integrated, in order to obtain solutions for the sought-after master integrals. 

Recently it has been noticed that the freedom of choosing different sets of master integrals can be exploited to find particular simple differential equations, the so called canonical forms~\cite{Henn:2013pwa}. Such forms are characterized by a total differential in $\dlog$ form and by the factorization of the dimensional regularization parameter from the kinematics. The arguments of the $\dlog$'s are called letters and together form the alphabet of our problem. While canonical forms greatly simplify the solution of differential equations and expose many of the underlying mathematical structures, finding these forms can be challenging and has been the subject of several publications~\cite{MullerStach:2012mp,Argeri:2014qva,Gehrmann:2014bfa,Lee:2014ioa,Meyer:2016slj,Dlapa:2020cwj,Henn:2020lye}. In this work we follow the algorithm outlined in~\cite{Argeri:2014qva,DiVita:2014pza} and first find a form that is linear in the dimensional regularization parameter, which then is brought to the $\eps$-factorized form through the Magnus algorithm~\cite{Argeri:2014qva}. This approach has been shown to work in a variety of applications~\cite{DiVita:2014pza,Bonciani:2016ypc,DiVita:2017xlr,Mastrolia:2017pfy,DiVita:2018nnh,Primo:2018zby,DiVita:2019lpl}, including cases with many kinematic invariants, non-planar integrals and non-rational alphabets. The resulting $\eps$-factorized form is then expanded in the ratio of lepton over Z boson mass, leading to a rational alphabet. This new alphabet agrees with the one presented in the massless calculation~\cite{Bonciani:2016ypc,vonManteuffel:2017myy} up to the desired logarithms in the lepton mass. Due to the rational nature of the alphabet our results can be conveniently written in terms of generalized polylogarithms up to weight four. The results and the corresponding rotation matrix found by the Magnus algorithm are given in the ancillary files of the \texttt{arXiv} version of this publication.

Throughout this computation we made use of publicly available codes {\tt Kira}~\cite{Maierhoefer2018a} and {\tt Reduze}~\cite{Manteuffel2012a} for the generation of the IBPs and the differential equations, {\tt LiteRed}~\cite{Lee:2013mka} for the dimensional reduction identities, {\tt SecDec}~\cite{Borowka:2015mxa} for the numerical validation of our results and {\tt GiNaC}~\cite{Vollinga:2004sn} for the numerical evaluation of the generalized polylogs.

%% file: tex/notation.tex
\section{Notation}
\label{sec:Notation}

\begin{figure}
\begin{center}
\subfloat[Topology 1]{\includegraphics[width = 1.6in]{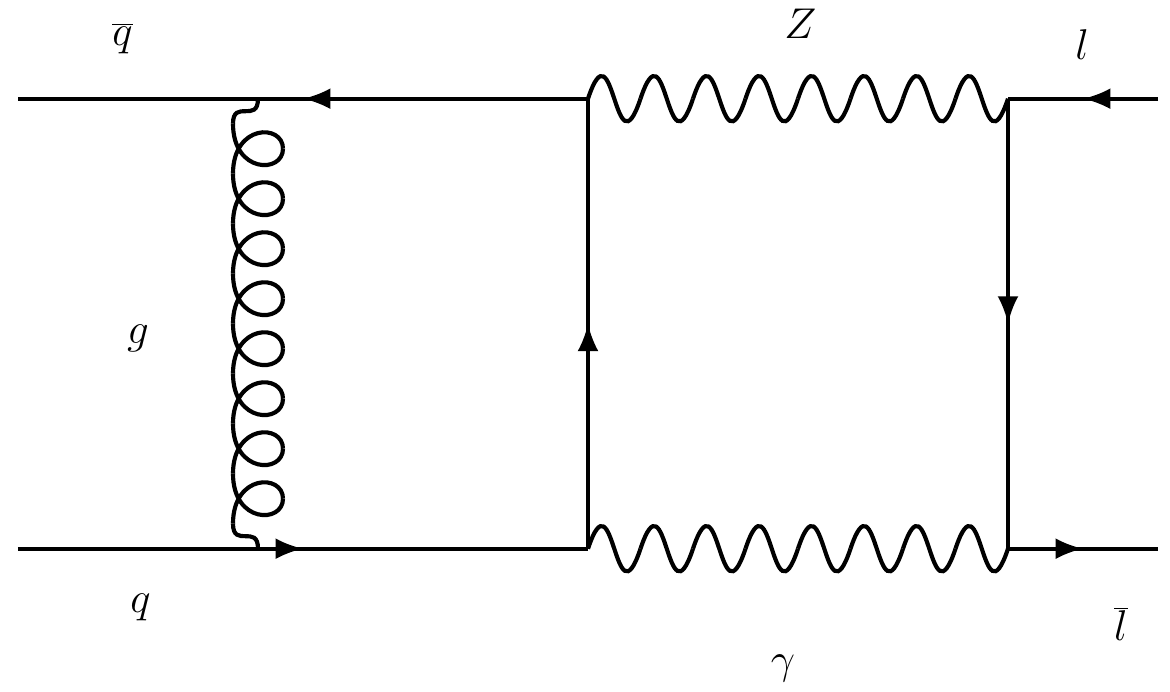}} \, \, \,
\subfloat[Topology 2]{\includegraphics[width = 1.6in]{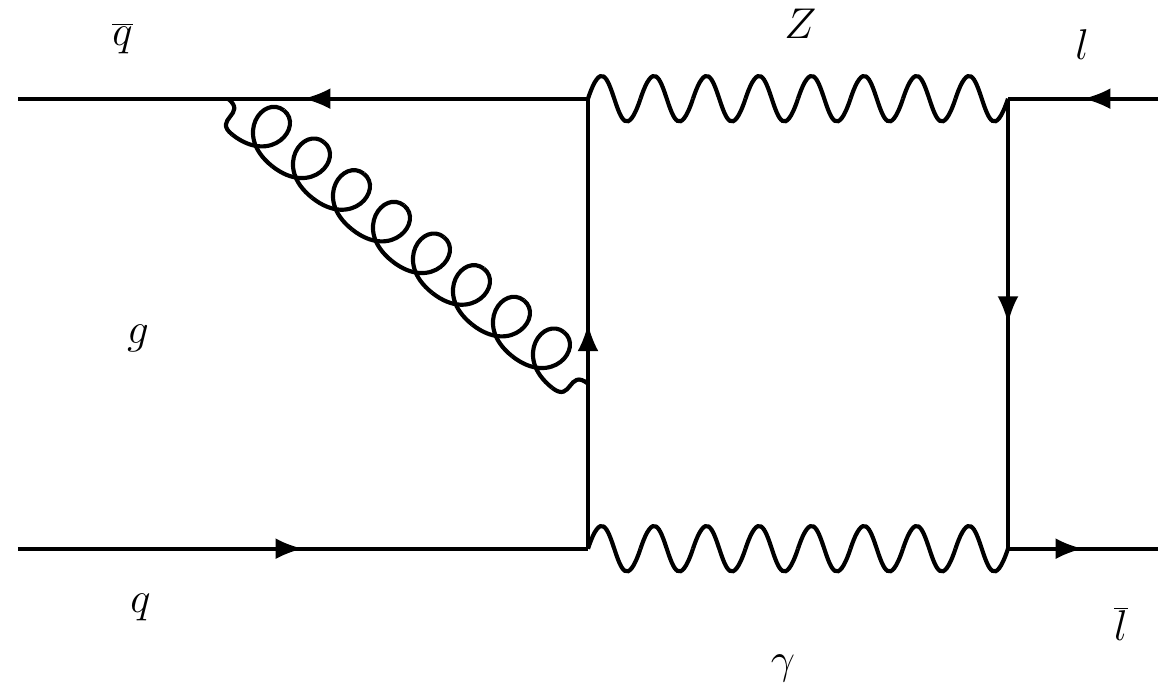}} \, \, \,
\subfloat[Topology 3]{\includegraphics[width = 1.6in]{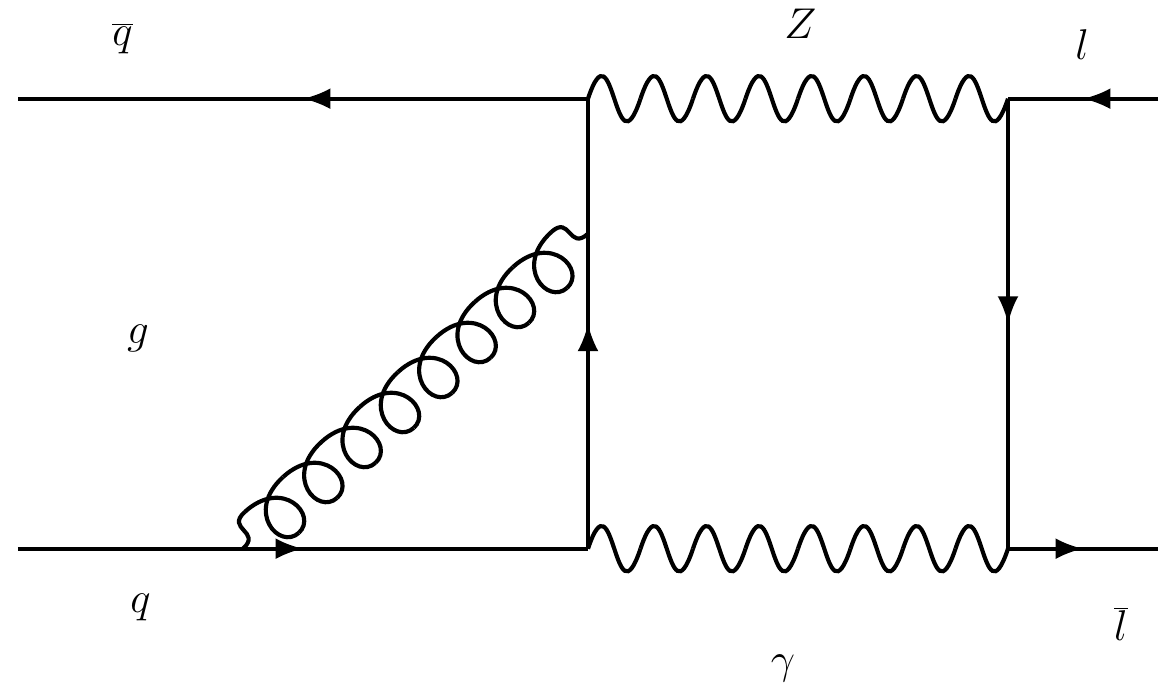}}
\caption{Example Feynman diagrams for each of the topologies cosidered for the non-factorizable mixed QCD-QED correction to the neutral current DY process.}
\label{fig:DY2L}
\end{center}
\end{figure}

This article considers the mixed QCD-QED two-loop corrections to the Drell-Yan process  
\begin{align}
  q(p_1) + \bar{q}(p_2)  \to    l^+(p_3) +  l^-(p_4)\,,
\end{align}
specified by the following kinematics
\begin{align}
p_1^2&=p_2^2=0\,,\quad p_3^2=p_4^2=m_l^2\,,\nn
  s = (p_1+p_2)^2\,,  \quad t&= (p_1-p_3)^2\,, \quad u=(p_2-p_3)^2=2m_l^2 -s-t  \, ,
\end{align}
where the lepton mass $m_l$ was kept non-zero throughout the calculation. In particular our work is concerned with the quantum corrections involving the exchange of a photon and a Z boson between the initial and final state, depicted in figure~\ref{fig:DY2L}. The appearing integrals are of the form
\begin{gather}
I(n_1,\ldots,n_9)
\equiv
  \frac{1}{C(\eps)}\int \dd^d k_1 \dd^d k_2\,
  \frac{1}{\Den_{1}^{n_1} \ldots \Den_{9}^{n_9}}\,,
\label{eq:def:ourintegrals}
\end{gather}
with the normalization factor
\begin{align}
  C(\eps) = {} & - \frac{1}{2} \left( \frac{1}{i \pi^{d/2}} \right)^2\, \left(\frac{m_z^2}{\mu^2}\right)^{2\epsilon} \,\Gamma\left(1+\epsilon\right) \,  \Gamma\left(1-\epsilon\right) \, \Gamma\left(1+2\epsilon\right) \,,
\label{eq:intmeasure1}
\end{align}
and the propagator definitions
\begin{gather}
\Den_1 = k_1^2,\quad
\Den_2 = k_2^2,\quad
\Den_3 = (k_1+p_1)^2,\quad
\Den_4 = (k_1+p_1+p_2)^2, \nn
\Den_5 = (k_2+p_1+p_2)^2-m_z^2,\quad
\Den_6 = (k_2+p_3)^2-m_l^2,\quad
\Den_7 = (k_1-k_2)^2, \nn
\Den_8 = (k_2+p_1)^2,\quad
\Den_9 = (k_1+p_3)^2\,. 
\label{eq:2Lfamily}
\end{gather}
In the above definitions $k_i$ denote the loop momenta and the normalization factor $C(\eps)$ has been chosen such that the canonical integral $\GG_{9}$ is set to one.

\section{Differential equations}
\label{sec:DEQ}
Integration-by-parts identities allow us to express the derivative of a complete set of Feynman integral in respect to some kinematic invariant as a linear combination of the initially chosen Feynman integrals. This leads to a coupled first order differential equation, which can be solved in order to determine those Feynman integrals.
For the process under consideration it is convenient to combine the invariants into three dimensionless ratios  
\begin{equation}
 -\frac{s}{m_z^2} = x, \qquad  -\frac{t}{m_z^2} = y,  \qquad
\frac{m_l^2}{m_z^2}=z .
  \label{eq:varsxyz}
\end{equation}
resulting in a set of three partial differential equations
\begin{equation}
\partial_x \FFvec =  \tilde{\AA}_x \FFvec, \quad  \partial_y \FFvec =  \tilde{\AA}_y \FFvec,  \quad \partial_z \FFvec = \tilde{\AA}_z \FFvec \, .
\label{eq:DEQ}
\end{equation}
The solution of these differential equations can be written as series of iterated integrals $W$ over the matrices $\tilde{\AA}$ and a vector of boundary constants $\FFvec_0$
\begin{align}
\FFvec=W \, \FFvec_0 \, .
\end{align}
By rescaling the master integrals $\FFvec$ with the appropriate powers in the dimensional regularization parameter $\eps=\frac{4-D}{2}$ we can ensure that the matrix $W$ exhibits a Taylor series in $\eps$ 
\begin{align}
    W=W^{(0)}+\eps \, W^{(1)} +\eps^2 \, W^{(2)}+\eps^3 \, W^{(3)}+\eps^4 \, W^{(4)} +\mathcal{O}(\eps^5) \, .
\end{align}
The matrices $W$ can be obtained by integrating one of the partial differential equations and then fixing the resulting integration constant by matching the derivative of the obtained solution successively to the other partial differential equations. In our case we choose to integrate first in $x$, then $y$ and finally $z$, resulting in the following recursive formulas
\begin{align}
    W^{(i)} & =W^{(i)}_x+W^{(i)}_y+W^{(i)}_z \, , \\
    W^{(i)}_x &= \int \tilde{\AA}_x W^{(i-1)} \, dx \, , \\
    W^{(i)}_y&= \int \tilde{\AA}_y W^{(i-1)}-  \partial_y W^{(i)}_x   \, dy  \, ,\\
    W^{(i)}_z&= \int \tilde{\AA}_z W^{(i-1)}-  \partial_z W^{(i)}_x-  \partial_z W^{(i)}_y   \,  dz \, ,
\end{align}
where at the first step $W^{(-1)}$ is replaced by the identity matrix. An important step in the integration of a differential equation is the identification of a functional basis that includes all integrals encountered during those integrations. For the presented integrals this was achieved by choosing a particular basis of master integrals, where the dimensional regularization parameter $\eps$ factorizes from the kinematics, which are encoded in a $\dlog$-form. After an expansion for small $z$ all arguments of the $\dlog$'s are simple rational functions, which enable us to express the integrals in terms of the well known generalized polylogarithms~\cite{Goncharov:polylog,Remiddi:1999ew,Gehrmann:2001jv,Vollinga:2004sn}
\begin{align}
    G(a;u)&=\int_0^u \frac{dt}{t-a} \, , \\
    G(a_n,\dots,a_1;u)&=\int_0^u \frac{dt}{t-a_n}G(a_{n-1},\dots,a_1;u)\, , \\
    G(\vec{0}_n;u)&=\frac{1}{n!}\log(u)^n \, ,
\end{align}
with weights $a_i$ and argument $u$. In our case the weights are drawn from the sets 
\begin{align}
    \left\{-1,0,-y,-\frac{y}{1+y} \right\}\, , \ \
    \left\{0,1 \right\}\, , \ \
    \left\{0 \right\} \, ,
\end{align}
for arguments $x,y$ and $z$ respectively. Due to the expansion for small $z$, some integrals can not be directly expressed as generalized polylogartihms. Nevertheless after repeatedly applying integration by parts identities of the form
\begin{align}
    \int_0^u h(t) \, G(\vec{a};t) \, dt = \Big[ H(t) \, G(\vec{a};t) \Big]_0^u - \int_0^u H(t) \,  \partial_t G(\vec{a};t) \, dt \, ,
\end{align}
we either obtain an integral corresponding to a generalized polylogarithm or a purely rational function.

%% file: tex/deq.tex
\section{$\eps$-factorized form}
As a first step towards the calculation of the relevant master integrals we identify a special set of master integrals, for which the dimensional regularization parameter factorizes from the kinematics. This can be achieved in a two step process through the Magnus algorithm, presented in~\cite{Argeri:2014qva,DiVita:2014pza}. First we identify a special set of master integrals
\begin{align*}
\FF_{1}&=\eps^2 \, \top{1}\,,  &
\FF_{2}&=\eps^2 \, \top{2}\,,  &
\FF_{3}&=\eps^2 \, \top{3}\,,  \\
\FF_{4}&=\eps^2 \, \top{4}\,,  &
\FF_{5}&=\eps^3 \, \top{5}\,,  &
\FF_{6}&=\eps^2 \, \top{6}\,,  \\
\FF_{7}&=\eps^2 \, \top{7}\,,  &
\FF_{8}&=\eps^2 \, \top{8}\,,  &
\FF_{9}&=(1-\eps)\eps^2 \, \top{9}\,,  \\
\FF_{10}&=\eps^3 \, \top{10}\,,  &
\FF_{11}&=\eps^3 \, \top{11}\,,  &
\FF_{12}&=\eps^2 \, \top{12}\,,  \\
\FF_{13}&=\eps^4 \, \top{13}\,,  &
\FF_{14}&=\eps^2 \, \top{14}\,,  &
\FF_{15}&=\eps^2 \, \top{15}\,,  \\
\FF_{16}&=\eps^2 \, \top{16}\,,  &
\FF_{17}&=\eps^3 \, \top{17}\,,  &
\FF_{18}&=\eps^2\, \top{18}\,,  \\
\FF_{19}&=\eps^2\, \top{19}\,,  &
\FF_{20}&=\eps^3\, \top{20}\,,  &
\FF_{21}&=\eps^3 \, \top{21}\,, \\
\FF_{22}&=(1-2\eps)\eps^3 \, \top{22}\,, &
\FF_{23}&=\eps^3 \, \top{23}\,, &
\FF_{24}&=\eps^3 \, \top{24}\,, \\
\FF_{25}&=\eps^3 \, \top{25}\,, &
\FF_{26}&=\eps^2 \, \top{26}\,, &
\FF_{27}&=\eps^4 \, \top{27}\,, \\
\FF_{28}&=\eps^3 \, \top{28}\,, &
\FF_{29}&=\eps^3 \, \top{29}\,, &
\FF_{30}&=\eps^2 \, \top{30}\,, \\
\FF_{31}&=\eps^3 \, \top{31}\,, &
\FF_{32}&=\eps^2 \, \top{32}\,, &
\FF_{33}&=\eps^4 \, \top{33}\,, \\
\FF_{34}&=\eps^3 \, \top{34}\,, &
\FF_{35}&=(1-2\eps)\eps^3 \, \top{35}\,, &
\FF_{36}&=\eps^3 \, \top{36}\,, \\
\FF_{37}&=(1-2\eps)\eps^2 \, \top{37}\,, &
\FF_{38}&=\eps^4 \, \top{38}\,, &
\FF_{39}&=\eps^3 \, \top{39}\,, \\
\FF_{40}&=\eps^2 \, \top{40}\,, &
\FF_{41}&=\eps^3 \, \top{41}\,, &
\FF_{42}&=\eps^3 \, \top{42}\,, \\
\FF_{43}&=\eps^4\, \top{43}\,, &
\FF_{44}&=\eps^3 \, \top{44}\,, &
\FF_{45}&=\eps^4 \, \top{45}\,, \\
\FF_{46}&=\eps^3\, \top{46}-\eps^3 \, \frac{t-m_l^2}{m_z^2} \, \top{47}\,, &
\FF_{47}&=\eps^3 \, \top{47}\,, &
\FF_{48}&=\eps^4 \, \top{48}\,, \\
\FF_{49}&=\eps^4\, \top{49}\,, &
\FF_{50}&=\eps^4 \, \top{50}\,, &
\FF_{51}&=\eps^4 \, \top{51}\,.
\stepcounter{equation}\tag{\theequation}
\label{def:LBasis}
\end{align*}
which are depicted in figure~\ref{fig:MIs}. These integrals satisfy a precanonical differential equation, which has a linear dependence on the dimensional regularization parameter $\eps$. Secondly we follow the Magnus algorithm to build a rotation matrix, which identifies the corresponding factorized master integrals
\begin{align*}
  \GG_{1}&= -s \,\FF_1\,, &
  \GG_{2}&= s^2 \,  \FF_2\,, \nn
  \GG_{3}&= -s\, \FF_3\,, &
  \GG_{4}&= -\frac{s \, r_2}{2 \,m_z} \left(\FF_1 - \FF_3 +2\,m_l^2 \FF_4 \right) \,, \nn
  \GG_{5}&= -s \, r_1 \,  \FF_5\,, &
  \GG_{6}&= - s \,  \FF_6\,,  \nn
  \GG_{7}&= - s \, \FF_7\,, &
  \GG_{8}&= 2m_z^2 \, \FF_7 +(m_z^2-s)\,\FF_{8}\,, \nn
  \GG_{9}&=  \FF_{9} \,,   &
  \GG_{10}&= -(t-m_l^2) \, \FF_{10}\,, \nn 
  \GG_{11}&= -s \, \FF_{11}\,, &
  \GG_{12}&= -\frac{m_z^2}{m_z^2+s}\left(  \frac{3 s}{2}\FF_{6} + \FF_{9} + (m_z^2-s) \, s \,  \FF_{12} \right)\,, \nn 
  \GG_{13}&= -s\, \FF_{13}\,,  &
  \GG_{14}&= m_l^2 \, \FF_{14}\,, \nn
  \GG_{15}&=-t \, \FF_{15}\,,  &
  \GG_{16}&= 2m_l^2 \,\FF_{15}- (t-m_l^2)\,\FF_{16}  \,, \nn 
 \GG_{17}&= r_1 \, \FF_{17}  \,, & 
  \GG_{18}&=  m_l^2 \, r_1 \,  \FF_{18}\,, \nn 
  \GG_{19}&= s \left(\frac{3}{2} \, \FF_{17}+m_l^2  \,  \FF_{18}-m_l^2  \, \FF_{19} \right)\,, \ \ \ \ \ \  &
  \GG_{20}&= r_1 \, \FF_{20}\,, \nn
 \GG_{21}&= s \, (t-m_l^2) \FF_{21} \,, & 
  \GG_{22}&= -\left(t-m_l^2\right) \FF_{22} , \nn
  \GG_{23}&= -(m_z^2-s)(t-m_l^2) \FF_{23} \,, &
  \GG_{24}&= -m_z^2(t-m_l^2)\, \FF_{24}\,,   \nn
  \GG_{25}&= s \, t \, \FF_{25} \,, &
  \GG_{26}&= -m_l^2 \, s \left( \, \FF_{25} - (t-m_l^2)\, \FF_{26} \right)\,,  \nn
  \GG_{27}&= (u-m_l^2) \FF_{27}  \,, &
  \GG_{28}&=  s \, (t-m_l^2) \, \FF_{28} \,, \nn
  \GG_{29}&= r_1 \,\FF_{29} \,, &
  \GG_{30}&= m_z \, r_2  \left(3 \,\FF_{29}+(m_z^2-s) \FF_{30} \right), \nn
  \GG_{31}&= -\left(t-m_l^2\right) \FF_{31} ,  && \nn 
  \GG_{32}&= \rlap{$ \displaystyle -\frac{r_2}{4m_z(m_l^2 +t)}\Big[ 12\, m_z^2 
  \, t\,\FF_{31}+\left(t-m_l^2\right){}^2 \left(2 \, \FF_{15}+\FF_{16}+6 \,\FF_{31}+4m_z^2 \, \FF_{32}\right) \Big. $} \nn 
   \rlap{$ \displaystyle \Big. +2 
   \left(t-m_l^2\right)\,\FF_9 +4 \,  m_z^4 \, t  \,\FF_{32}\Big]\, ,$} \nn
  \GG_{33}&= (u-m_l^2)  \, \FF_{33} \,,&
  \GG_{34}&= -(m_z^2-s)(t-m_l^2)\,\FF_{34}\,,  \nn
  \GG_{35}&= r_1 \, \FF_{35} \,, &
  \GG_{36}&= -(m_z^2-s)(t-m_l^2) \,\FF_{36}\,,  \nn
  \GG_{37}&= - \frac{m_z}{r_2}   \left( \frac{1}{2} \FF_{9} - m_l^2 \, \FF_{14} -m_l^2 \, \FF_{37} \right) \,, &
  \GG_{38}&= r_1\,\FF_{38}\,, \nn
  \GG_{39}&= -s \, r_1\,  \FF_{39} \,, && \nn
  \GG_{40}&= \rlap{$ \displaystyle s\left( \frac{1}{2}\FF_1-\frac{
   1}{2}\FF_3- \FF_{17} -\FF_{38} + m_l^2 \left(\FF_4 - \FF_{18}  \right)+ s\,\FF_{39}+\frac{1}{2}r_2^2 \,\FF_{41} \right) -m_z^2 \, \FF_{29}  $} && \nn
  \rlap{$ \displaystyle -\frac{ m_z^2\,}{r_2^2} \left( \frac{1}{2}\FF_9  -m_l^2 \, \FF_{14}- m_l^2\, \FF_{37} \right)+\frac{1}{2}  \left(s-m_z^2\right) \left( \,\FF_{29} + m_z^2\,\FF_{30}+2 \,s \, m_l^2 \,\FF_{40} \right) \, ,  $} && \nn
  \GG_{41}&=  m_z \, r_1 \, r_2 \, \FF_{41} \,,&
  \GG_{42}&=  m_z^2 \, r_1  \, \FF_{42}\,,  \nn
  \GG_{43}&= -\left(t-m_l^2\right) \, \FF_{43} \,, &
  \GG_{44}&= -m_z^2 \left(t-m_l^2\right) \left(\FF_{21}+ (m_z^2- s) \FF_{44}\right) \,, \nn
  \GG_{45}&= s \left(t-m_l^2\right) \FF_{45} \,,&
  \GG_{46}&= -s \, m_z \, r_2 \left(2 F_{45}+m_z^2 \,\FF_{46}\right) \,,  \nn
  \GG_{47}&= \rlap{$ \displaystyle  -\frac{1}{2} s \left(2(t- m_l^2)+m_z^2\right)\left( m_z^2 \,\FF_{46}+2 \,\FF_{45} +s\, t \,\FF_{47}\right) - s \left(m_l^4-t^2\right)\FF_{47}$} && \nn
  \GG_{48}&= \rlap{$ \displaystyle \sqrt{m_l^4
   \left(m_z^2-s\right)^2-2 m_l^2
   \left(t \left(m_z^2-s\right)^2+m_z^2 \,  s\left(m_z^2+s\right)\right)+\left(s\,t-m_z^2 (s+t)\right)^2} \,  \FF_{48}\,, $}  && \nn
  \GG_{49}&= -s \left(t-m_l^2\right) \left(s-m_z^2\right)\,\FF_{49} \,,&
  \GG_{50}&=-s  \,r_1 \, \FF_{50}\,,  \nn
  \GG_{51}&= \rlap{$ \displaystyle  \left(t-m_l^2\right) \left(s-m_z^2\right)\FF_{48} -s \left(s-m_z^2\right) \left(m_l^2 \, \FF_{49} -\FF_{51} \right) \,, $} && 
\label{def:CanonicalBasis}
\stepcounter{equation}\tag{\theequation}
\end{align*}
where we introduced the abbreviations $r_1=\sqrt{-s}\sqrt{4m_l^2-s}$ and $r_2=\sqrt{m_z^2-4m_l^2}$. 
The transformations given in eq. \eqref{def:CanonicalBasis} and \eqref{def:LBasis} are provided in the ancillary files accompanying the \texttt{arXiv} version of this publication. The presented integrals satisfy a set of three partial differential equations
\begin{equation}
\partial_x \GGvec = \eps \AA_x \GGvec, \quad  \partial_y \GGvec = \eps \AA_y \GGvec,  \quad \partial_z \GGvec = \eps \AA_z \GGvec \, ,
\label{eq:partialcanonicalDEQ}
\end{equation}
where the dimensional regularization parameter $\eps$ has been factorized from the kinematic dependence encoded in $\AA$.

\include{tex/deq-figure_MIs}

\section{Solution}

The vastly different size of the lepton mass compared to the Z boson mass allows us to expand the differential equations for small values of $z=\frac{m_l^2}{m_z^2}$. This expansion captures the important logarithms in the lepton mass, while keeping the analytic complexity at a manageable level. In particular for the leading term in the $z$-expansion the matrices $\AA_x$ and $\AA_y$ are independent of $z$ and are completely comprised of rational entries. Combining the expanded differential equations into a total differential 
\begin{equation}
d \GGvec = \eps \dA \GGvec \, ,
\label{eq:canonicalDEQ}
\end{equation}
 exposes the $\dlog$ form of our problem
\begin{equation}
\dA = \sum_{i=1}^{7} \MM_i  \dlog(\eta_i) \, ,
\end{equation}
with an alphabet that up to the logarithm in the lepton mass is identical to the massless case~\cite{Bonciani:2016ypc,vonManteuffel:2017myy}
 \begin{align}
 \begin{alignedat}{3}
\eta_1 & =1+x\,,&\quad
\eta_2 & =x\,, & \quad 
\eta_3 & =y\,, \nn
\eta_4 & =z\,, & \quad
\eta_5 & =1-y\,,&\quad
\eta_6 & =x+y\,,\nn
\eta_7 & =x+y+x \,y  \,.&
\end{alignedat} \stepcounter{equation}\tag{\theequation}
\label{alphabet}
 \end{align}
A precise approximation of the integrals appearing in the amplitude is only guaranteed if the integrals $\top{i}$ are well described within the expansion. For this reason the kinematic factors defined in eq.~\eqref{def:CanonicalBasis} need to be considered, when determining the appropriate point at which we can drop all higher order terms. Expanding the kinematic factors for small lepton masses (small $z$) we find divergences in those factors
\begin{align}
\top{i} = \sum_j \frac{R^{-1}_{ij}}{z} \GG_{j}+\sum_j R^{0}_{ij}\GG_{j}+\mathcal{O}\left(z \right) \, .
\end{align}
For this reason we have expanded our canonical integrals $\GG_{j}$ up to the first order in $z$ such that all finite pieces are captured for the integrals $\top{i}$. These higher order terms can be integrated as described in section~\ref{sec:DEQ} and lead to rational terms in our canonical master integrals. Since the rational functions do not contribute to the alphabet and are irrelevant for the analytic continuation, the differential equations were integrated under the same constraints as the massless case
\begin{equation}
  \label{eq:positivityx}
  x>0\land 0<y<1 \land z>0   \,,
 \end{equation} 
which correspond to the unphysical region 
\begin{equation}
  s<0\,\land\,t<0\,\land\,m^2_l>0\,\land\,m_z^2>0 \, .
\end{equation}
Furthermore the analytic continuation to the physical region follows the same recipe that was already laid out in~\cite{Bonciani:2016ypc}.

\subsection{Boundary conditions}
By the nature of determining the Feynman integrals through their derivative, a boundary constant has to be specified in order to determine the exact solution. These constants can be obtained by either matching our solutions to easier integrals through some appropriate limit or by demanding the absence of unphysical thresholds in our alphabet. For our problem at hand we used the following conditions
\begin{itemize}
\item The integrals $\GG_{3,14}$ were provided as an independent input.
\item In the limit $m_l^2 \rightarrow0$, integrals $\GG_{37,43}$ were matched against their massless counterparts presented in~\cite{Bonciani:2016ypc}.
\item The boundary constants of integrals $\GG_{2,4,5,7,8,11,13,22,29,30,32,35,38,39,40,41}$  were fixed by demanding regularity in the limit $s\rightarrow 0$.
\item The integrals $\GG_{1,6,9,10,12,15,\dots,20}$ were matched against the full solutions presented in \cite{Mastrolia:2017pfy}.
\item Demanding regularity at the pseudo threshold $s=-t$, fixed the boundary constants of integrals $\GG_{21,23,25,\dots,28,33,34,36,44,\dots,47,50,51}$.
\item Taking the limit $t\rightarrow -m_z^2 $ and demanding its regularity results in relations between boundary constants of integrals $ \GG_{24,31,49}$. 
\item Integrals $\GG_{42,48}$ were fixed by demanding regularity in the limit $t \rightarrow \frac{-m_z^2\,s}{m_z^2-s}$. 
\end{itemize}
In our normalization~\eqref{eq:intmeasure1} the independent input integrals have the following expressions
\begin{align}
    \GG_{3}&= \left(x\,z\right)^{-\eps}\left(-2+6\, \zeta(2)\,\eps^2-\frac{15}{2}\, \zeta(4)\, \eps^4 +\mathcal{O}(\eps^5) \right) \, , \\
    \GG_{14}&=z^{-2\eps}\left( \frac{1}{2}+\zeta(2)\,\eps^2+5\, \zeta(3)\, \eps^3+\frac{45}{2}\zeta(4)\,\eps^4+\mathcal{O}(\eps^5) \right) \, .
\end{align}
The weight structure of the boundary constants established through the input integrals persists for all master integrals. Namely the boundary constants at order $\eps^i$ for $0\leq i \leq 4$ are proportional to the corresponding constants $1,0,\zeta(2),\zeta(3)$ and $\zeta(4)$.
The obtained boundary constants finalize the determination of the master integrals, whose expression are attached to the \texttt{arXiv} version of this publication.

\subsection{Consistency checks}

In order to verify the obtained solutions we performed a number of checks:
\begin{itemize}
\item We independently computed the equivalent one-loop integrals to our process and checked that all factorizable integrals $\GG_{1,2,3,4,5}$ analytically agree with the actual product of one-loop integrals.
\item Our expanded expressions were compared with the exact integrals obtained numerically by the package {\tt SecDec} and we found agreement within the expected errors due to our approximation. 
\item Following the steps outlined in section 7.1 of~\cite{Mastrolia:2017pfy}, we took the lepton mass to zero for 31\footnote{Only 29 zero lepton mass limits provide a true consistency check, since we already used two for the boundary fixing of integrals $\GG_{37}$ and $\GG_{43}$.} combinations of our integrals and found agreement with the solutions presented in \cite{Bonciani:2016ypc}.
\end{itemize}
The zero lepton mass limit for the last check was taken by performing a Jordan decomposition of the pole matrix $\AA_{z,-1}$ defined as
\begin{equation}
\AA_z=\frac{\AA_{z,-1}}{z}+\mathcal{O}[z^0] \, .
\end{equation}
The Eigenvectors $v=\sum{c_i \GG_i}$ of this matrix define linear combinations of our canonical integrals, whose behavior in the limit $z$ going to zero is defined by the corresponding Eigenvalue
\begin{align}
\lim_{z \rightarrow 0} v_i = z^{\lambda_i \eps} h(\eps)  \, ,
\end{align}
where $h(\eps)$ is some function that is independent of $z$.
The 31 Eigenvectors belonging to the Eigenvalue $0$ define linear combinations of our integrals that are finite in the limit $z\rightarrow 0$. These finite combinations can be employed in two ways. Firstly we can now safely take the $z\rightarrow 0$ limit directly on the analytic expressions of these combinations. Secondly we take the lepton mass to zero directly at the integrand level of these combinations. The resulting integrals can then be expressed in terms of the master integrals presented in the massless calculation~\cite{Bonciani:2016ypc}. Finally we found that the 31 expressions derived from our analytic results were in full agreement with the equivalent expressions derived by taking the lepton mass to zero at the integrand level. 

%% file: tex/deq-figure_MIs.tex
\begin{figure}
  \centering
  \captionsetup[subfigure]{labelformat=empty}
  \subfloat[$\mathcal{T}_1$]{%
    \includegraphics[width=0.105\textwidth]{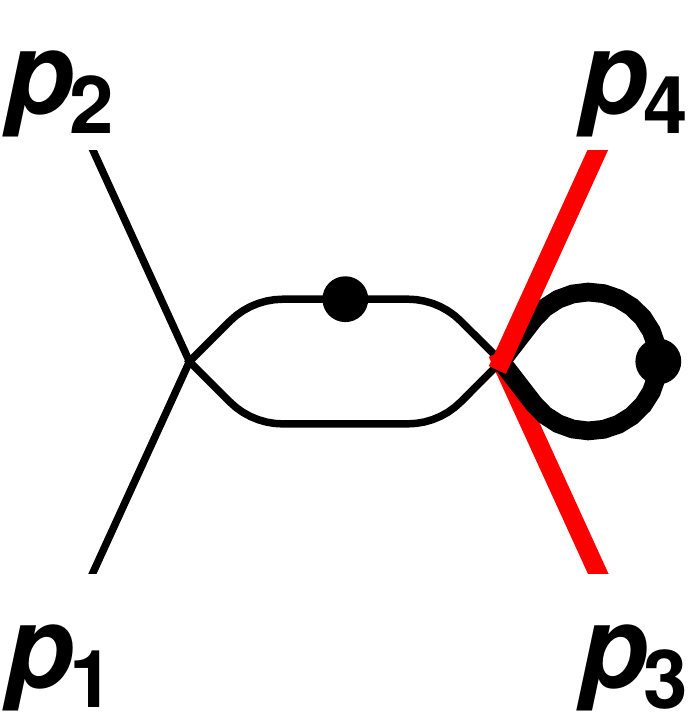}
  } \,
  \subfloat[$\mathcal{T}_2$]{%
    \includegraphics[width=0.105\textwidth]{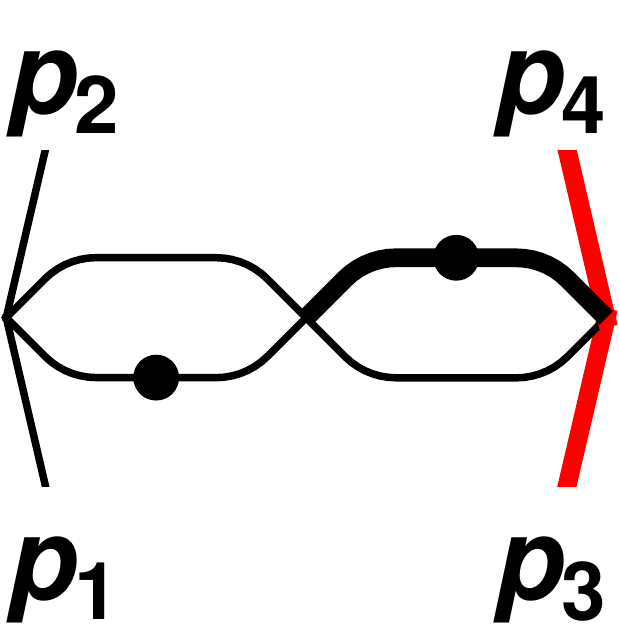}
  }\,
  \subfloat[$\mathcal{T}_3$]{%
    \includegraphics[width=0.105\textwidth]{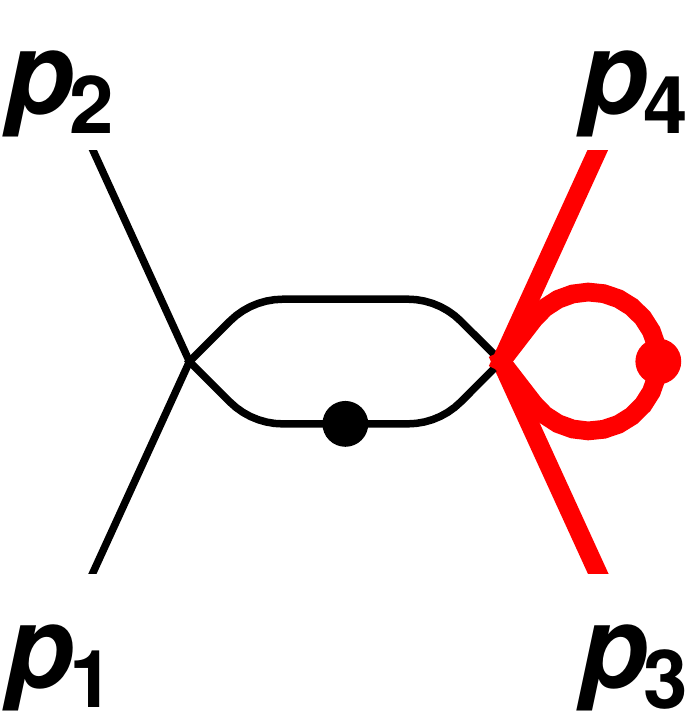}
  }\,
  \subfloat[$\mathcal{T}_4$]{%
    \includegraphics[width=0.105\textwidth]{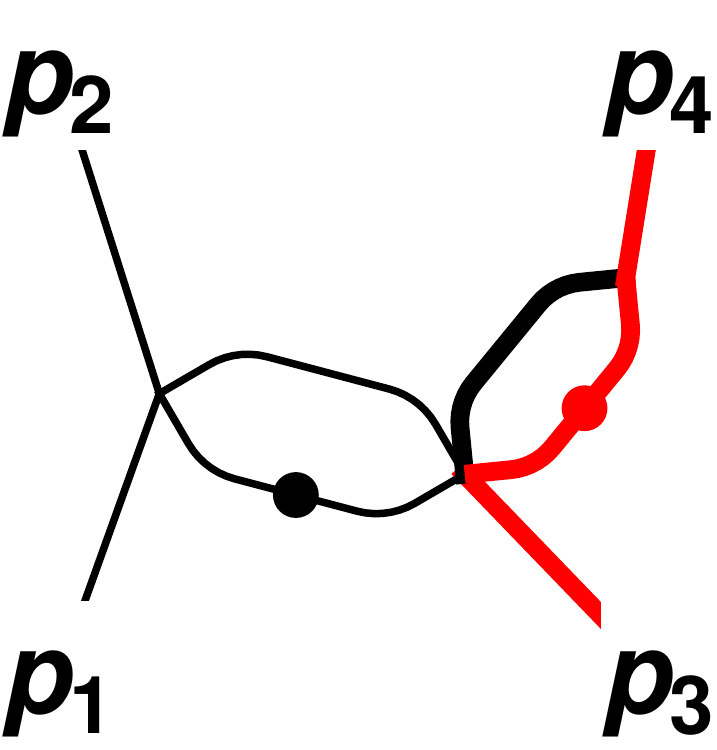}
  }\,
  \subfloat[$\mathcal{T}_5$]{%
    \includegraphics[width=0.105\textwidth]{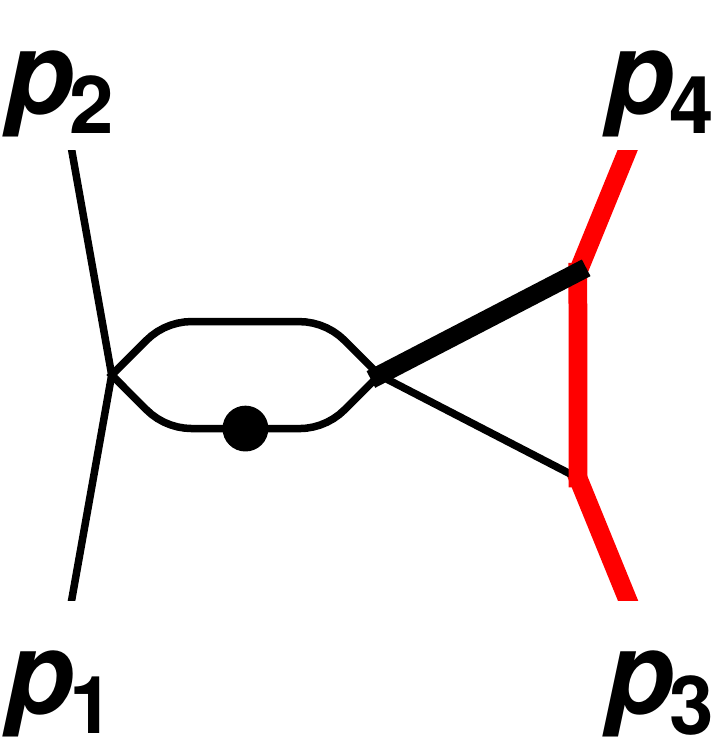}
  }
  \subfloat[$\mathcal{T}_6$]{%
    \includegraphics[width=0.105\textwidth]{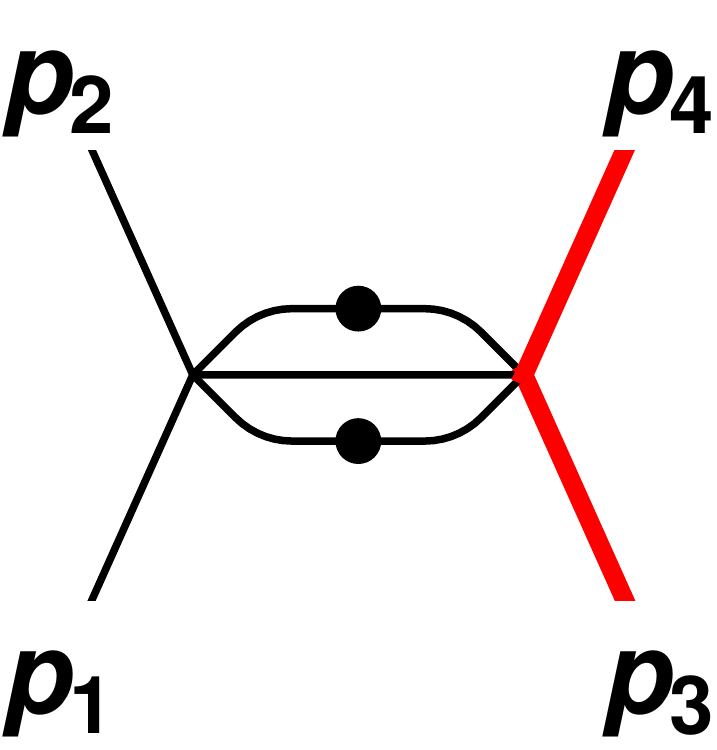}
  }
  \subfloat[$\mathcal{T}_7$]{%
    \includegraphics[width=0.105\textwidth]{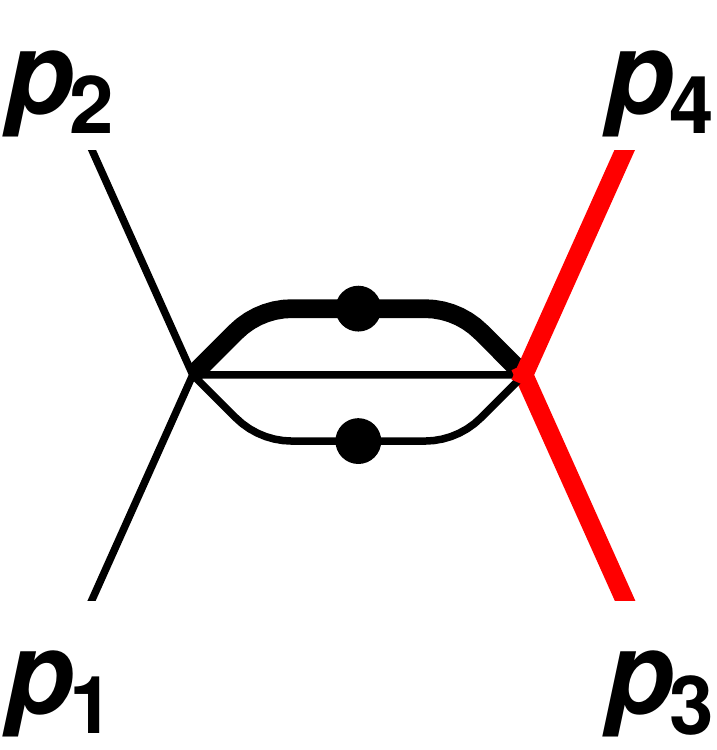}
  }\,
  \subfloat[$\mathcal{T}_8$]{%
    \includegraphics[width=0.105\textwidth]{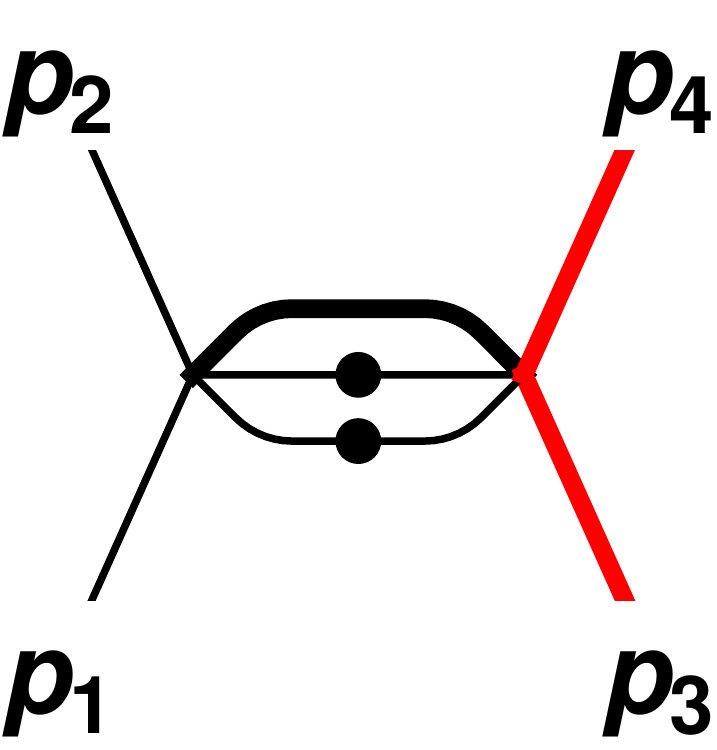}
  }\\
  \subfloat[$\mathcal{T}_9$]{%
    \includegraphics[width=0.105\textwidth]{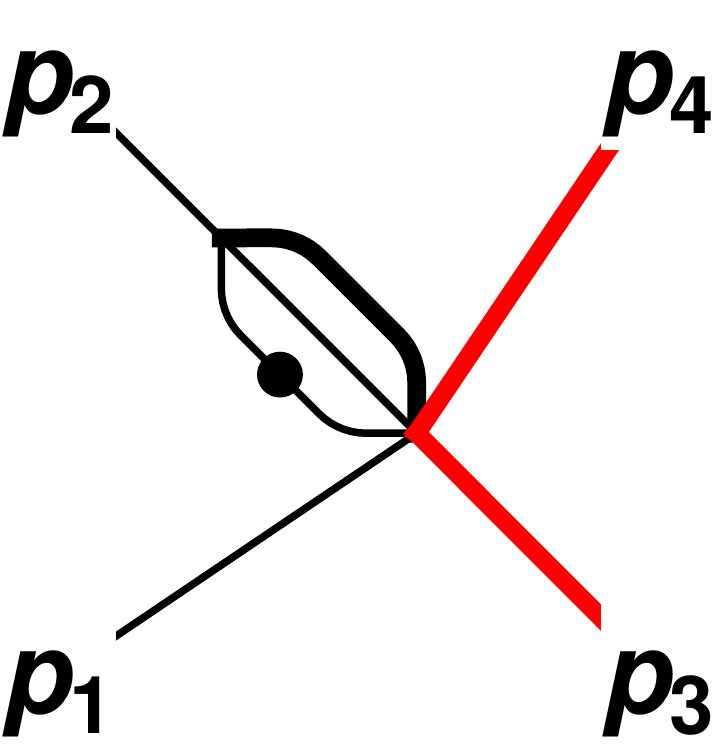}
  } \,
  \subfloat[$\mathcal{T}_{10}$]{%
    \includegraphics[width=0.105\textwidth]{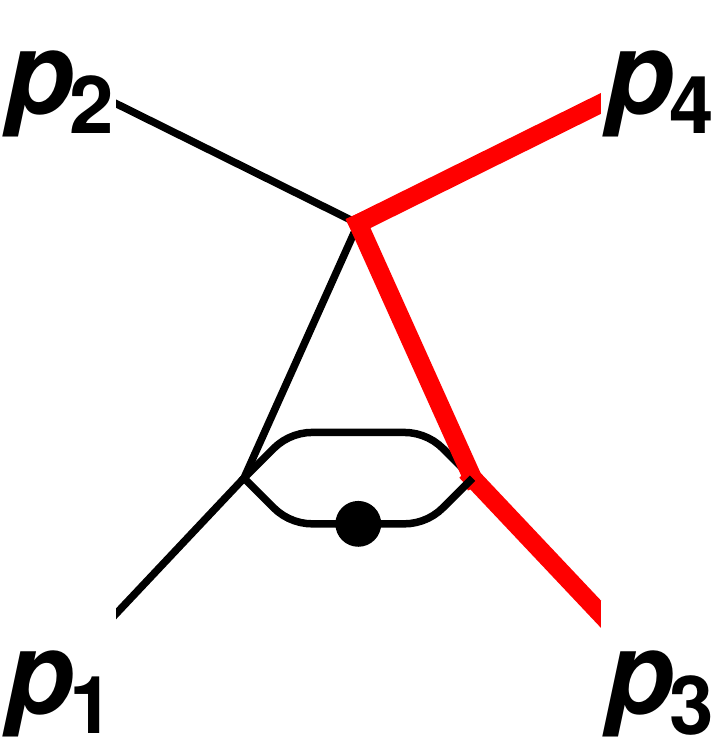}
  }\,
  \subfloat[$\mathcal{T}_{11}$]{%
    \includegraphics[width=0.105\textwidth]{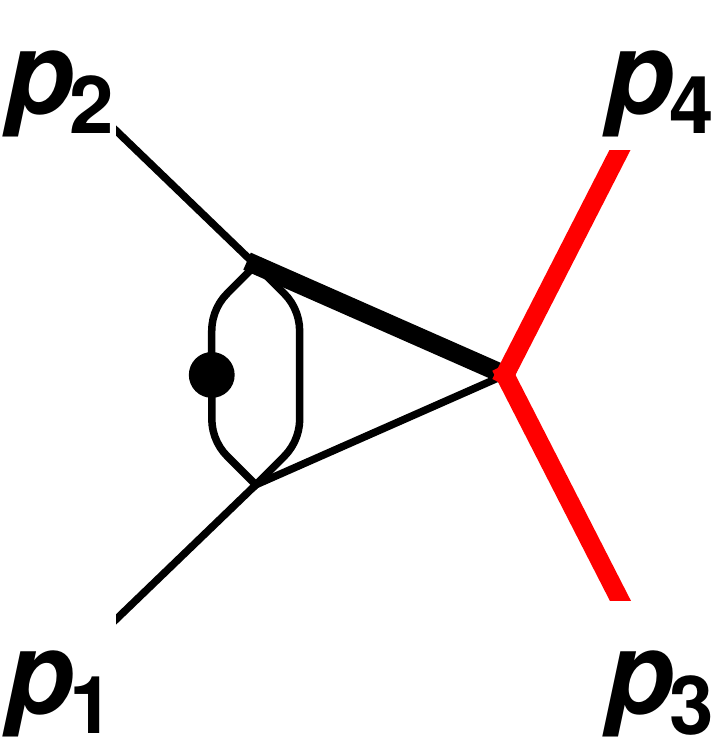}
  }\,
  \subfloat[$\mathcal{T}_{12}$]{%
    \includegraphics[width=0.105\textwidth]{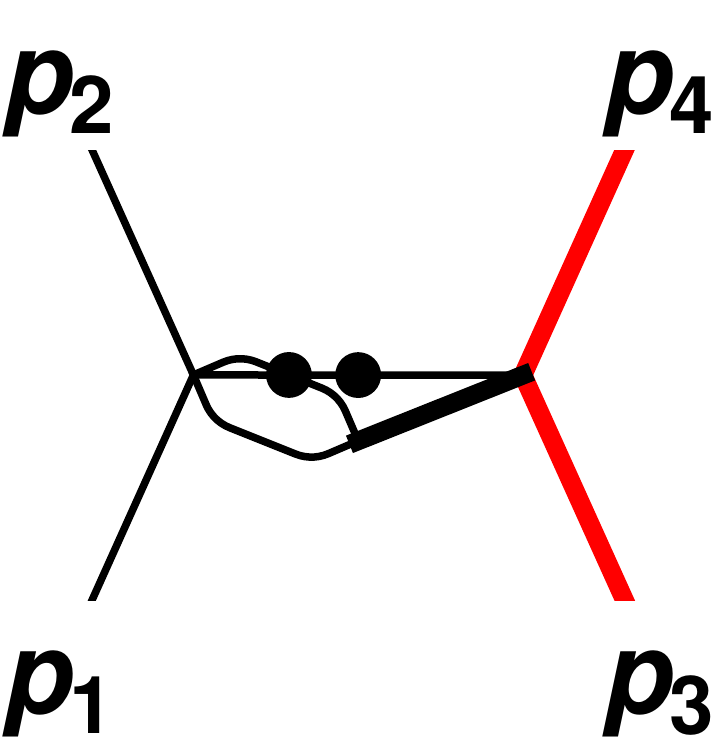}
  }\,
  \subfloat[$\mathcal{T}_{13}$]{%
    \includegraphics[width=0.105\textwidth]{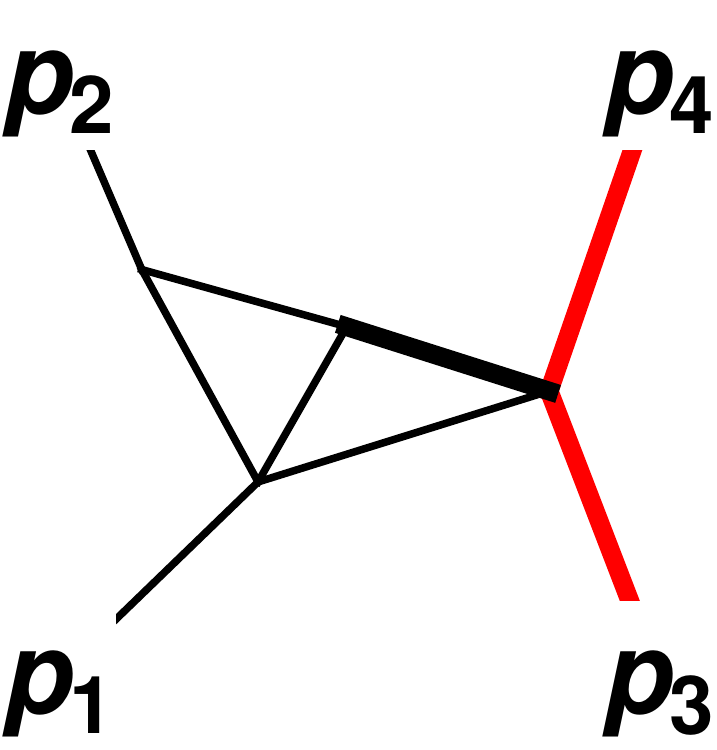}
  }\,
  \subfloat[$\mathcal{T}_{14}$]{%
    \includegraphics[width=0.105\textwidth]{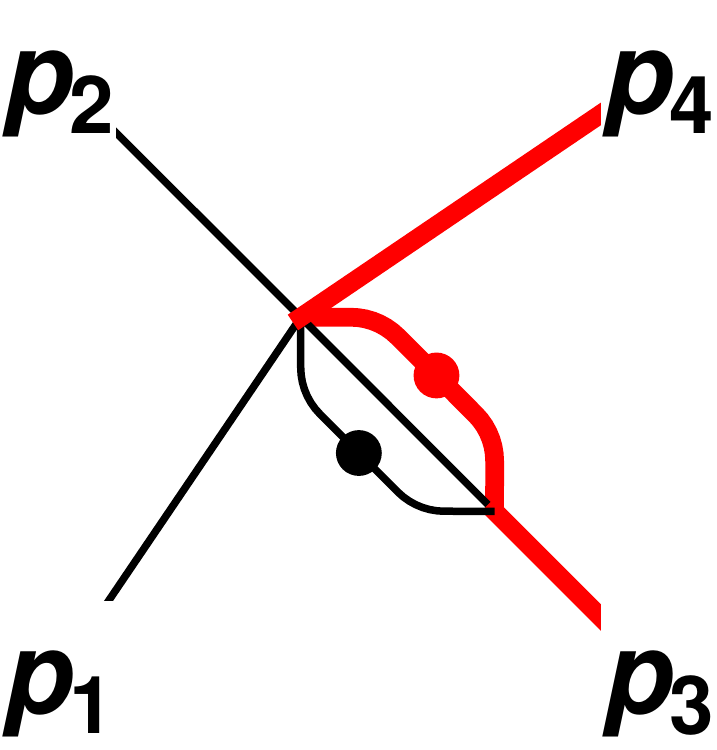}
  }\,
  \subfloat[$\mathcal{T}_{15}$]{%
    \includegraphics[width=0.105\textwidth]{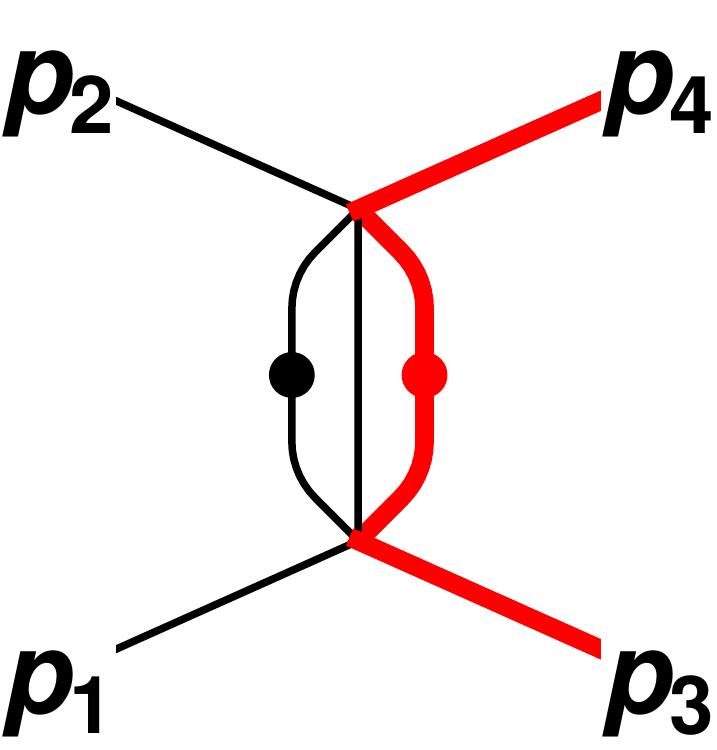}
  }\,
  \subfloat[$\mathcal{T}_{16}$]{%
    \includegraphics[width=0.105\textwidth]{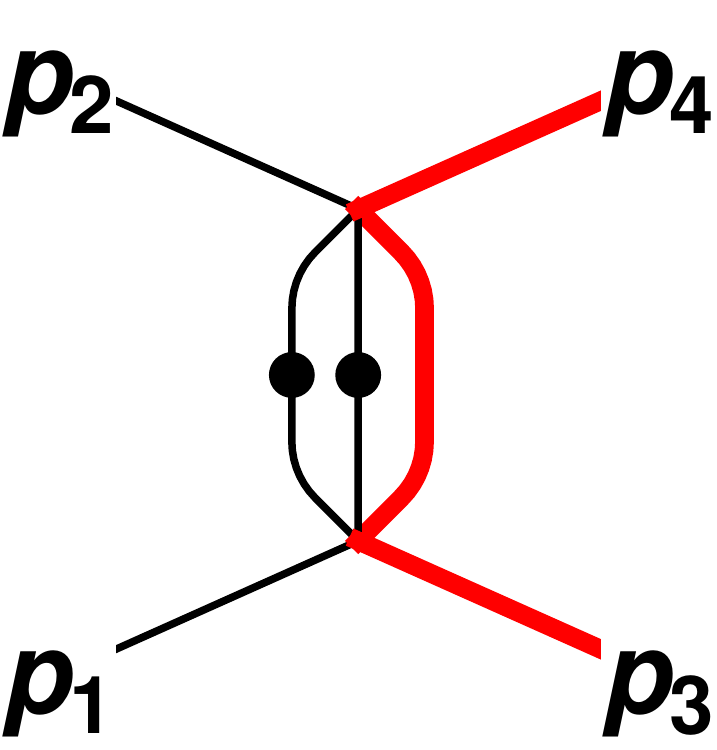}
  }\\
  \subfloat[$\mathcal{T}_{17}$]{%
    \includegraphics[width=0.105\textwidth]{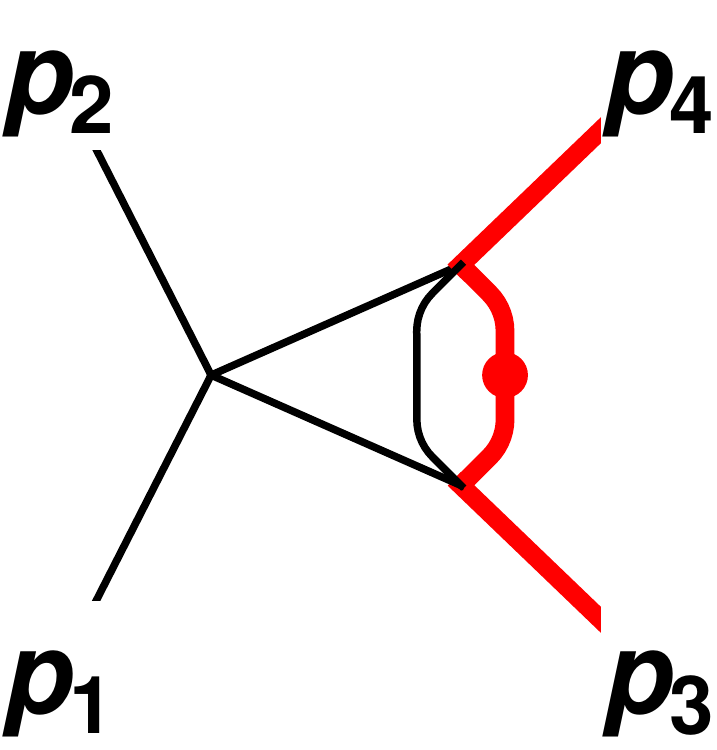}
  }\,
  \subfloat[$\mathcal{T}_{18}$]{%
    \includegraphics[width=0.105\textwidth]{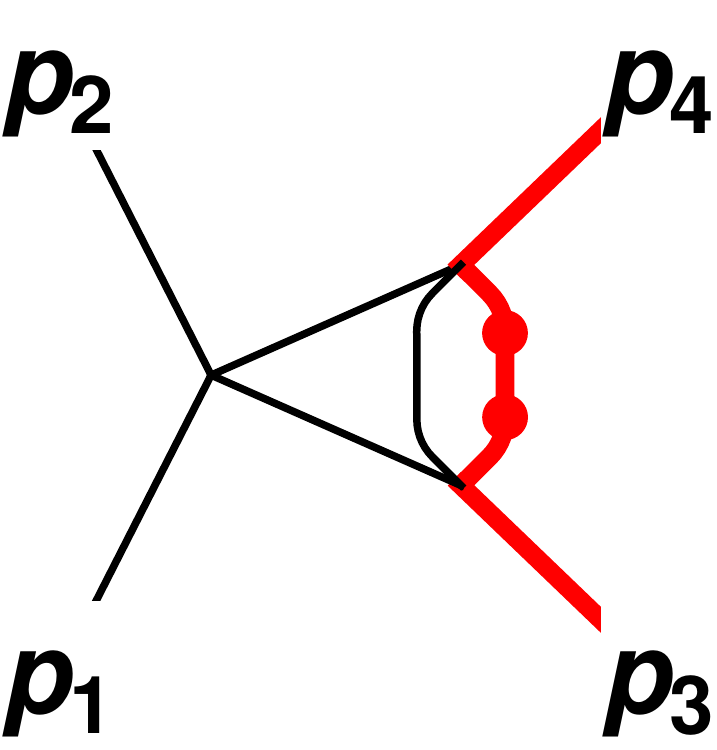}
  }\,
    \subfloat[$\mathcal{T}_{19}$]{%
    \includegraphics[width=0.105\textwidth]{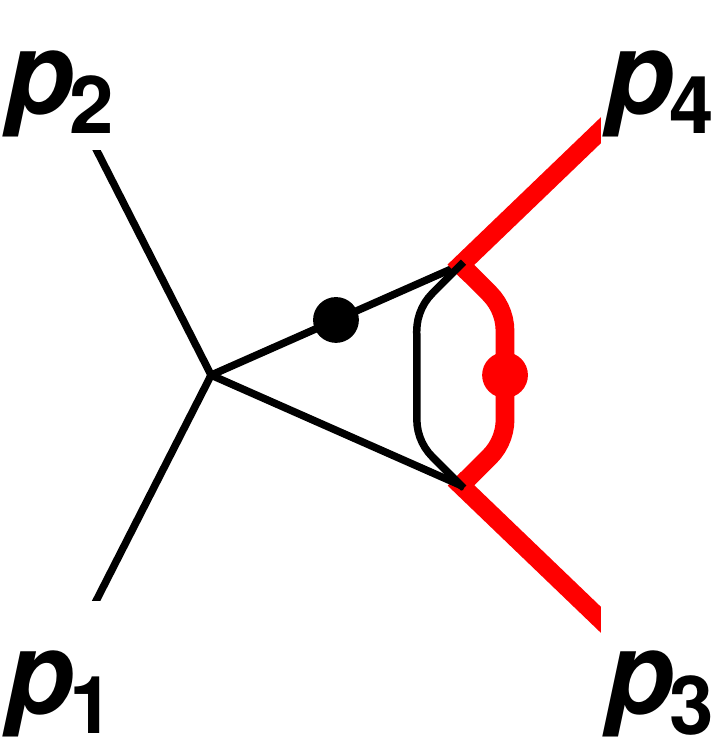}
  }\,
  \subfloat[$\mathcal{T}_{20}$]{%
    \includegraphics[width=0.105\textwidth]{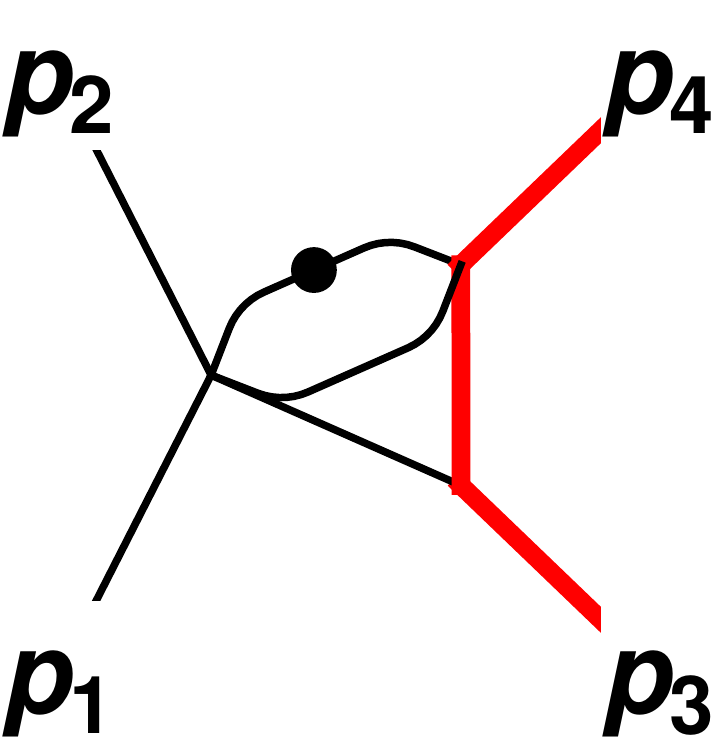}
  }\,
  \subfloat[$\mathcal{T}_{21}$]{%
    \includegraphics[width=0.105\textwidth]{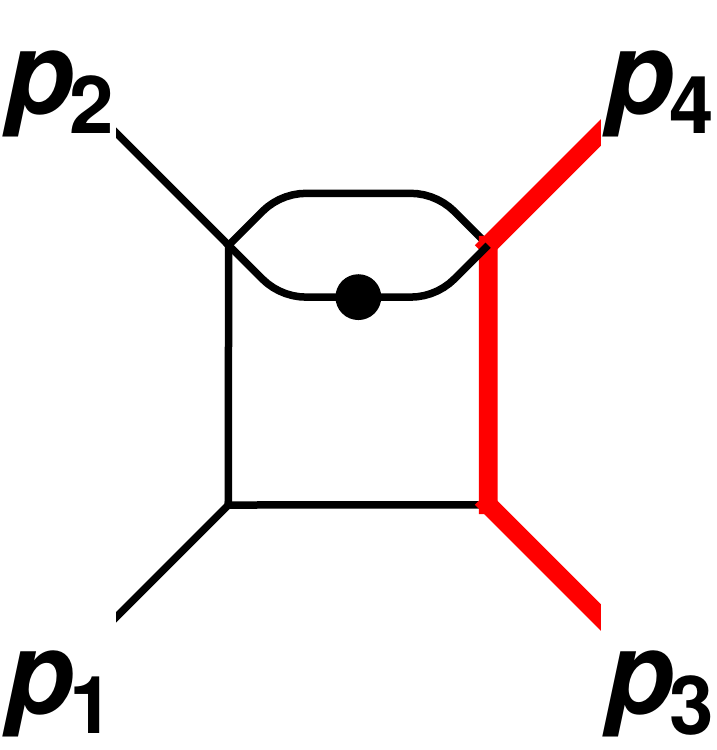}
  }\,
  \subfloat[$\mathcal{T}_{22}$]{%
    \includegraphics[width=0.105\textwidth]{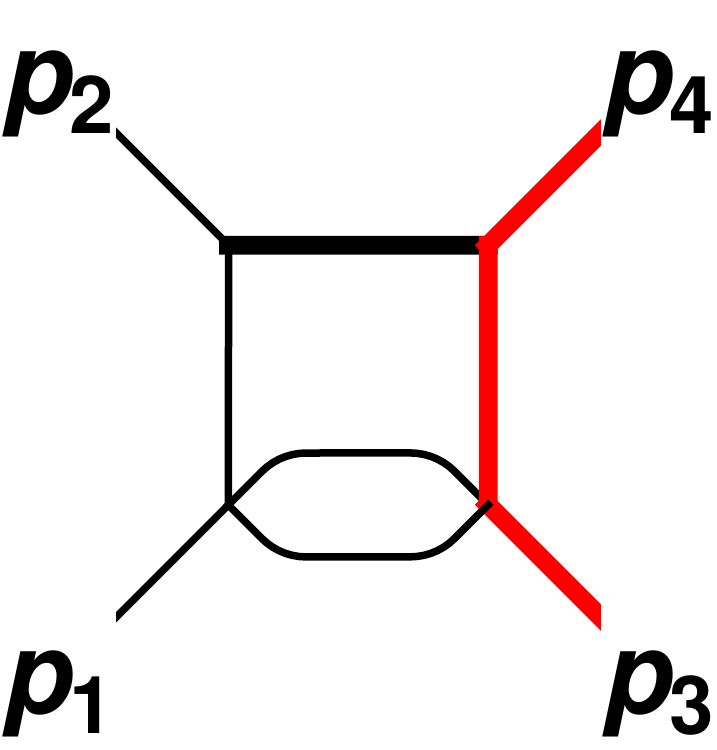}
  }\,
  \subfloat[$\mathcal{T}_{23}$]{%
    \includegraphics[width=0.105\textwidth]{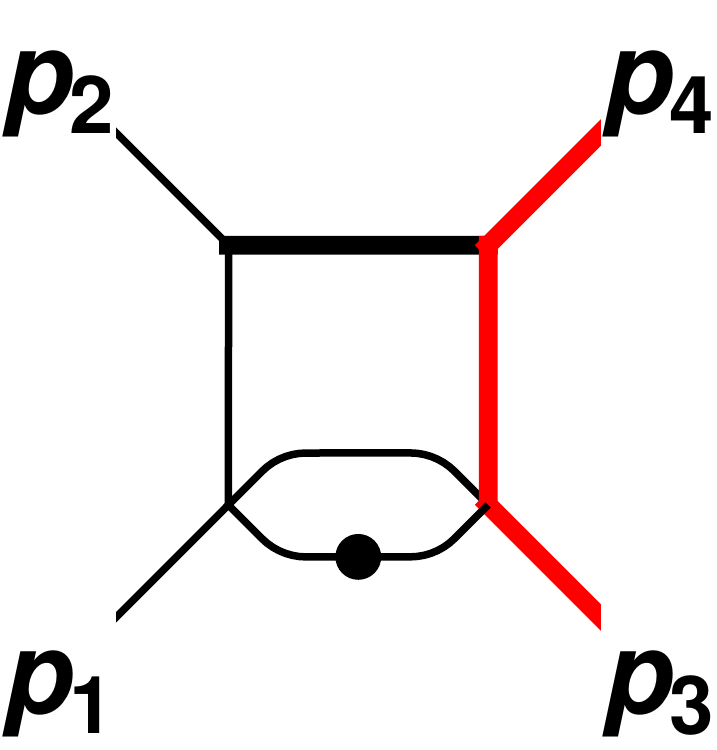}
  }\,
  \subfloat[$\mathcal{T}_{24}$]{%
    \includegraphics[width=0.105\textwidth]{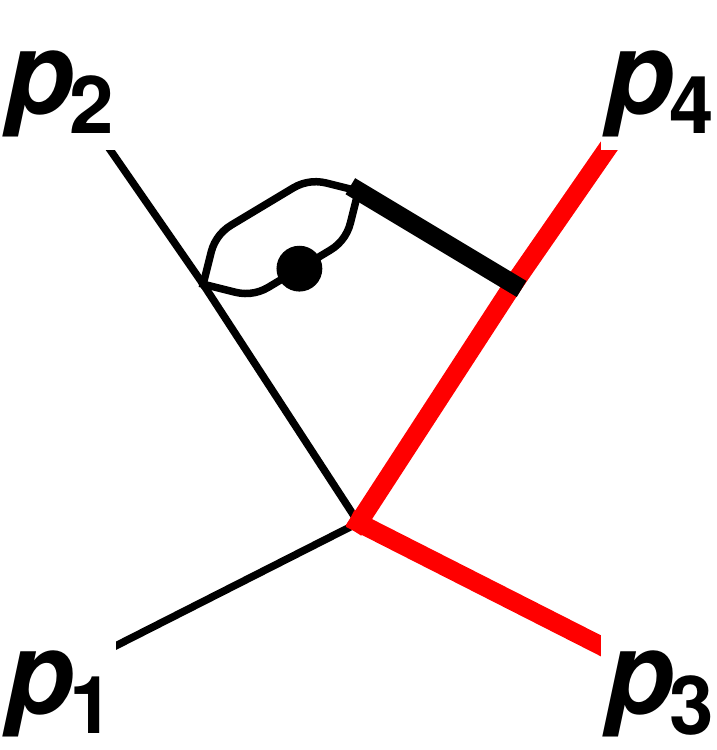}
  } \\
    \subfloat[$\mathcal{T}_{25}$]{%
    \includegraphics[width=0.105\textwidth]{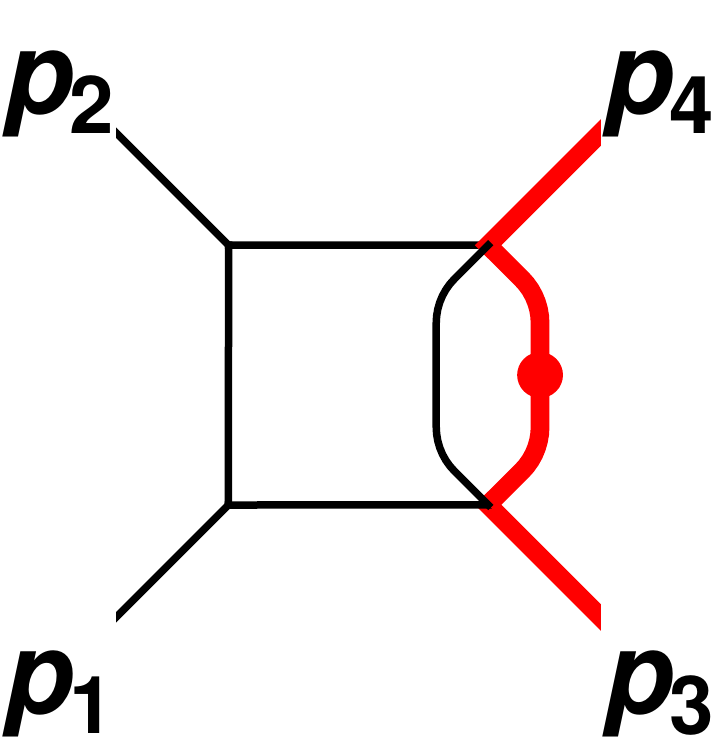}
  }\,
  \subfloat[$\mathcal{T}_{26}$]{%
    \includegraphics[width=0.105\textwidth]{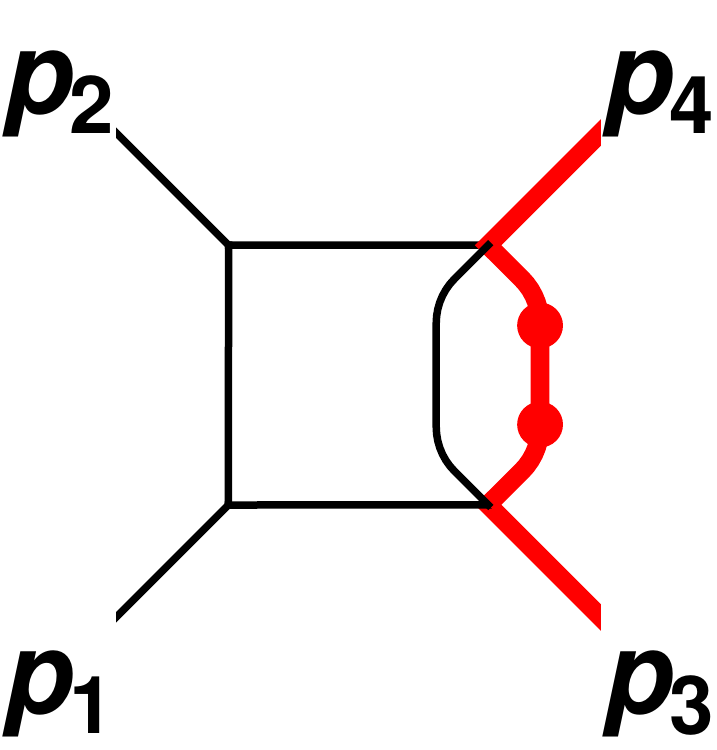}
  }\,
  \subfloat[$\mathcal{T}_{27}$]{%
    \includegraphics[width=0.105\textwidth]{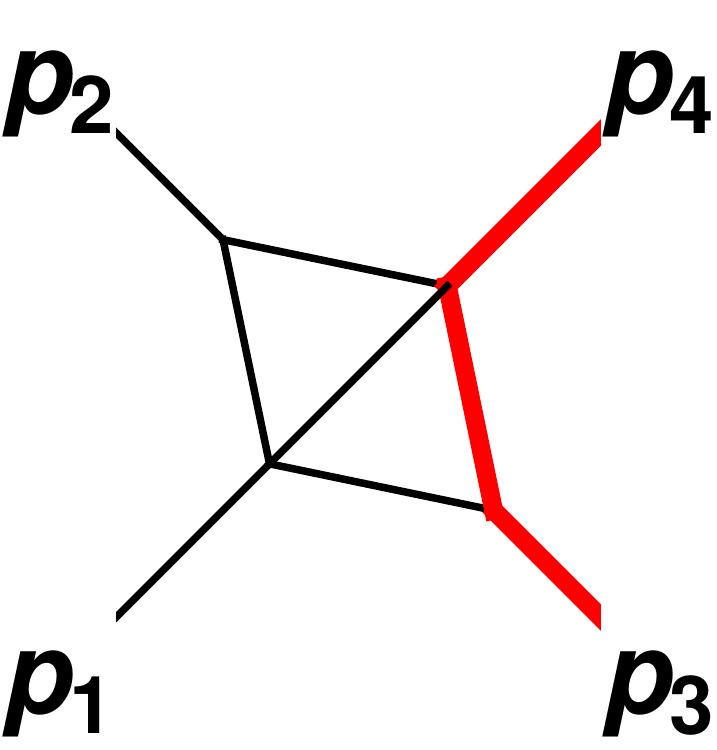}
  }\,
  \subfloat[$\mathcal{T}_{28}$]{%
    \includegraphics[width=0.105\textwidth]{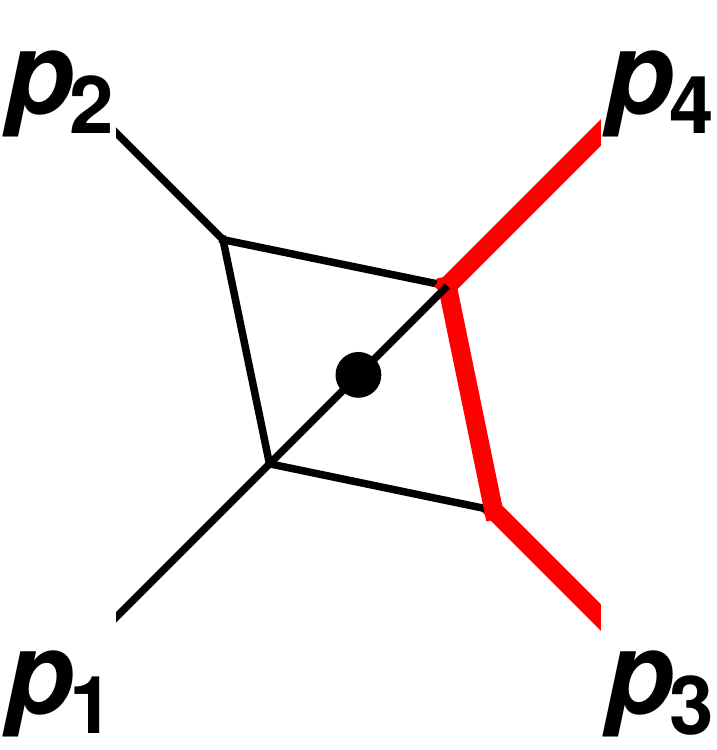}
  }\,
  \subfloat[$\mathcal{T}_{29}$]{%
    \includegraphics[width=0.105\textwidth]{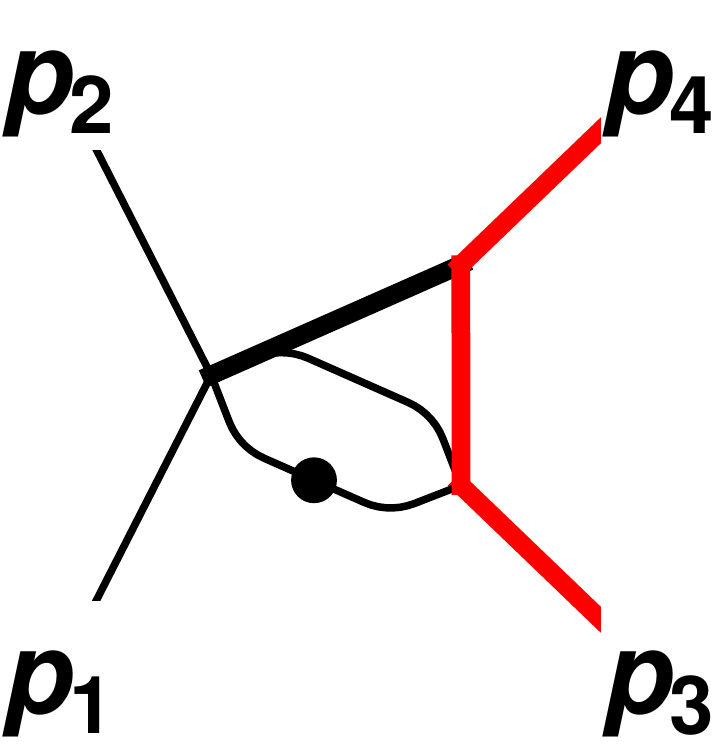}
  }\,
  \subfloat[$\mathcal{T}_{30}$]{%
    \includegraphics[width=0.105\textwidth]{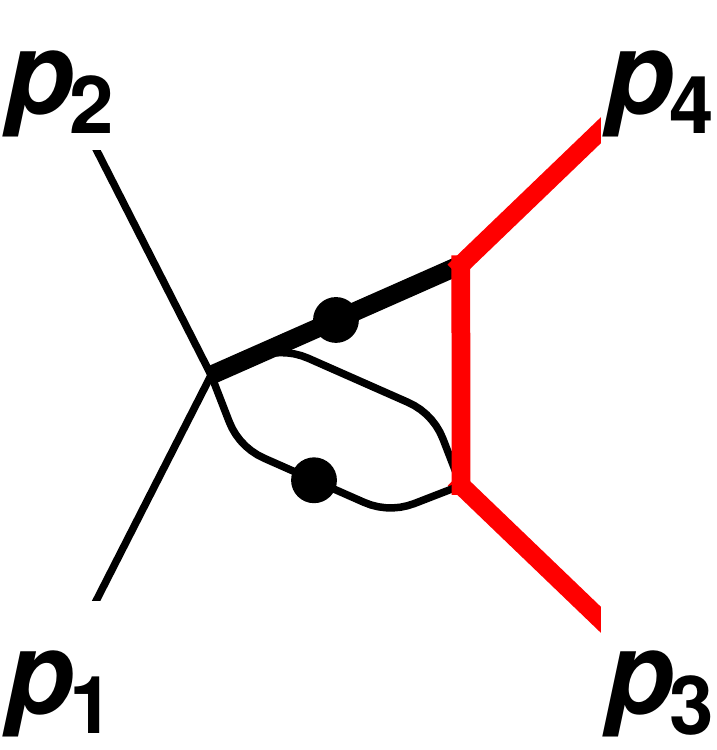}
  }\,
      \subfloat[$\mathcal{T}_{31}$]{%
    \includegraphics[width=0.105\textwidth]{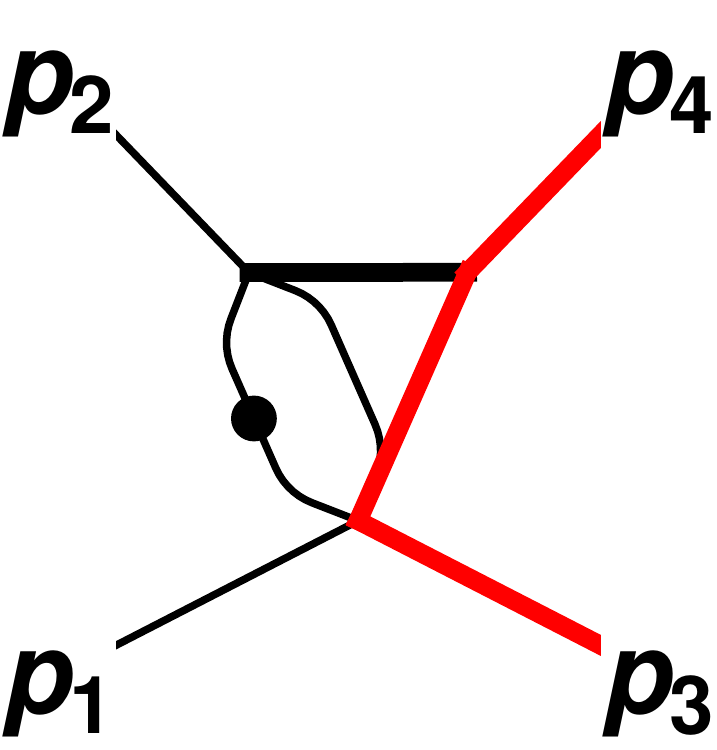}
  }\,
  \subfloat[$\mathcal{T}_{32}$]{%
    \includegraphics[width=0.105\textwidth]{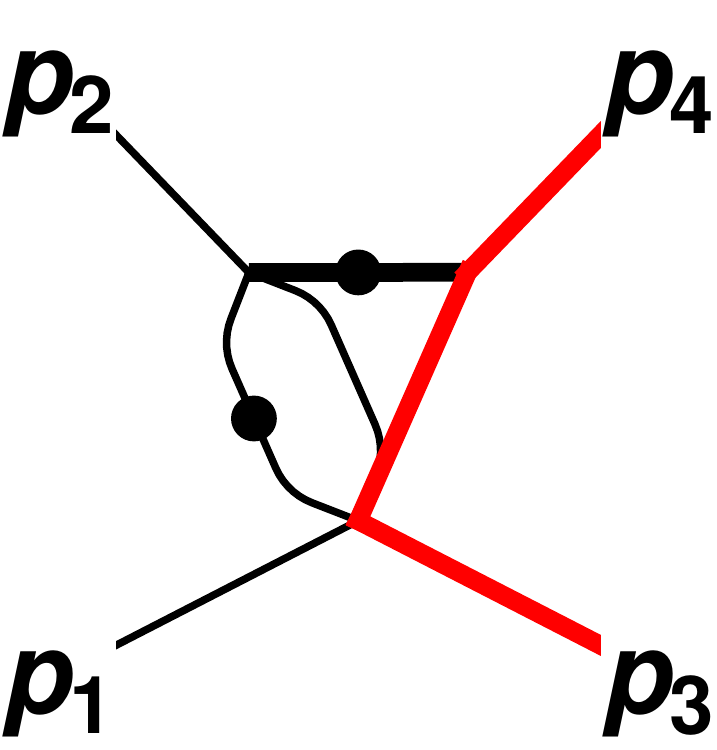}
  }\\
  \subfloat[$\mathcal{T}_{33}$]{%
    \includegraphics[width=0.105\textwidth]{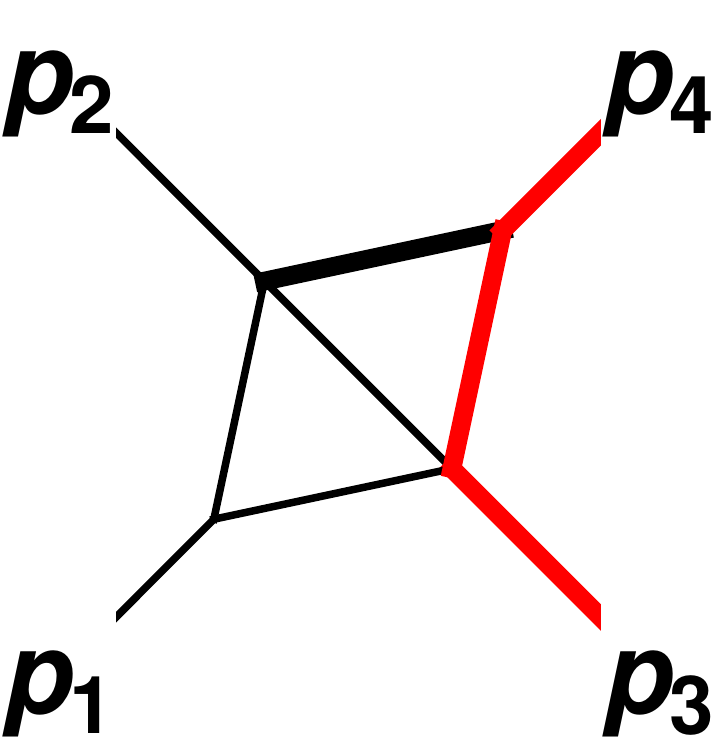}
  }\,
  \subfloat[$\mathcal{T}_{34}$]{%
    \includegraphics[width=0.105\textwidth]{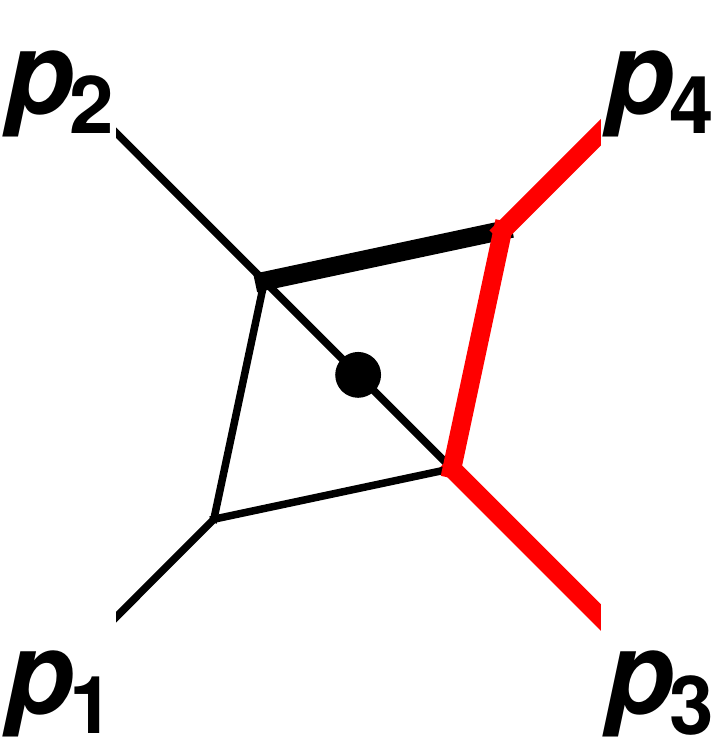}
  }\,
  \subfloat[$\mathcal{T}_{35}$]{%
    \includegraphics[width=0.105\textwidth]{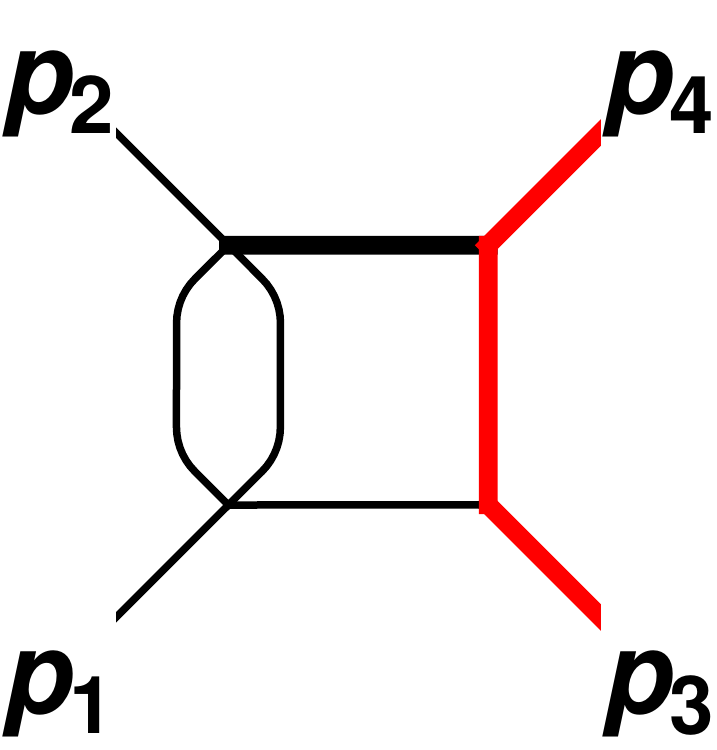}
  }\,
  \subfloat[$\mathcal{T}_{36}$]{%
    \includegraphics[width=0.105\textwidth]{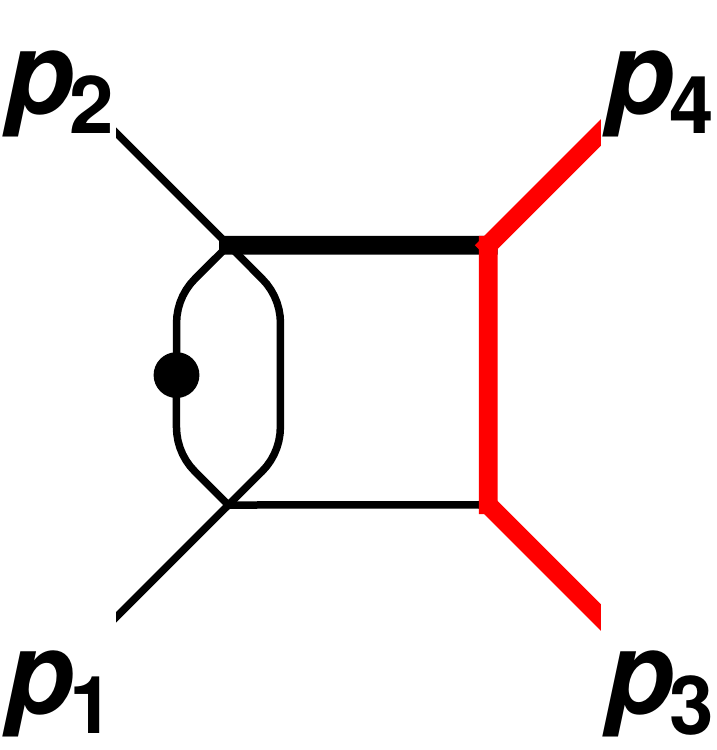}
  }\,
      \subfloat[$\mathcal{T}_{37}$]{%
    \includegraphics[width=0.105\textwidth]{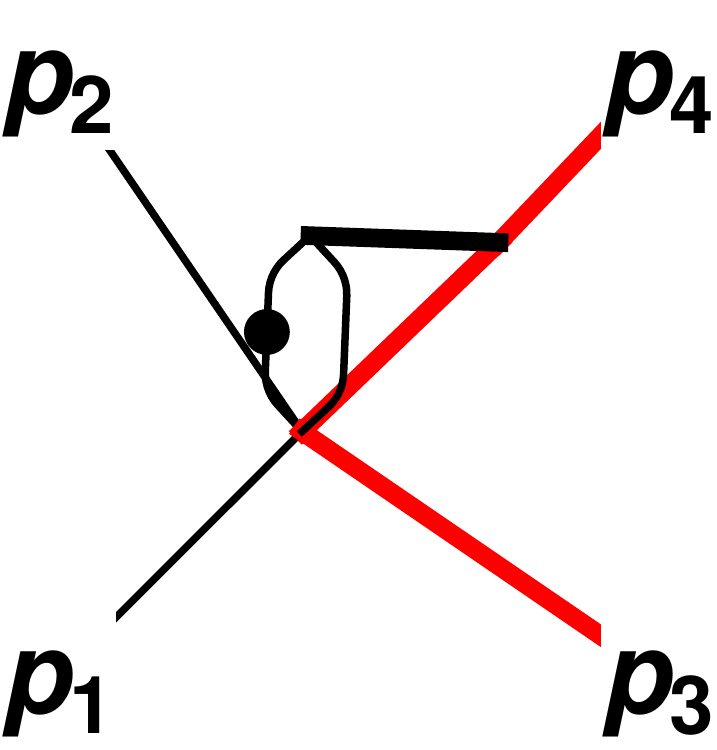}
  }\,
  \subfloat[$\mathcal{T}_{38}$]{%
    \includegraphics[width=0.105\textwidth]{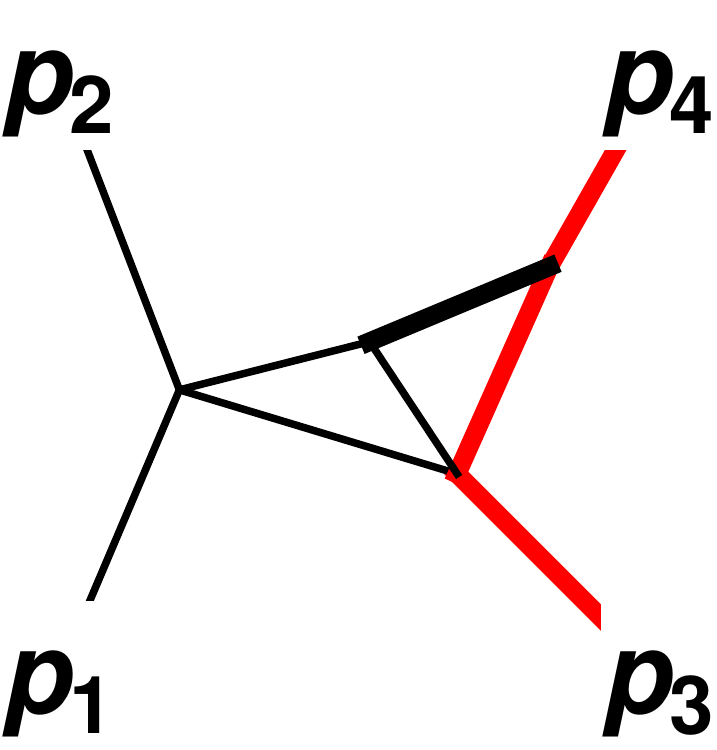}
  }\,
  \subfloat[$\mathcal{T}_{39}$]{%
    \includegraphics[width=0.105\textwidth]{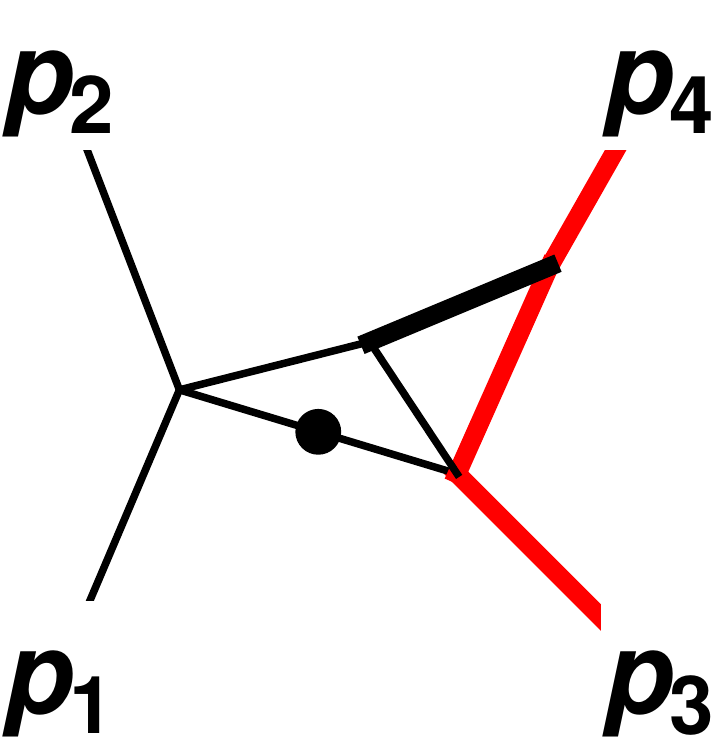}
  }\,
  \subfloat[$\mathcal{T}_{40}$]{%
    \includegraphics[width=0.105\textwidth]{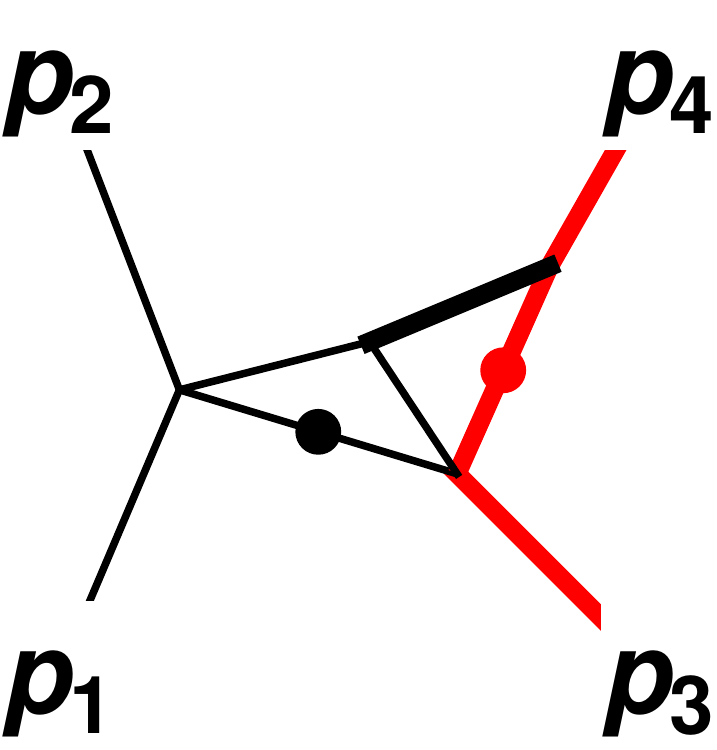}
  }\\
  \subfloat[$\mathcal{T}_{41}$]{%
    \includegraphics[width=0.105\textwidth]{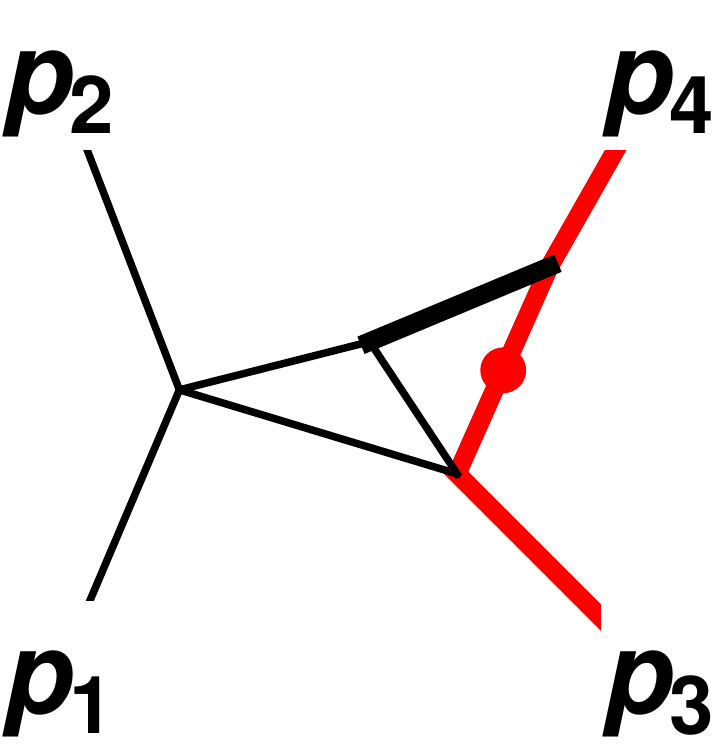}
  }\,
  \subfloat[$\mathcal{T}_{42}$]{%
    \includegraphics[width=0.105\textwidth]{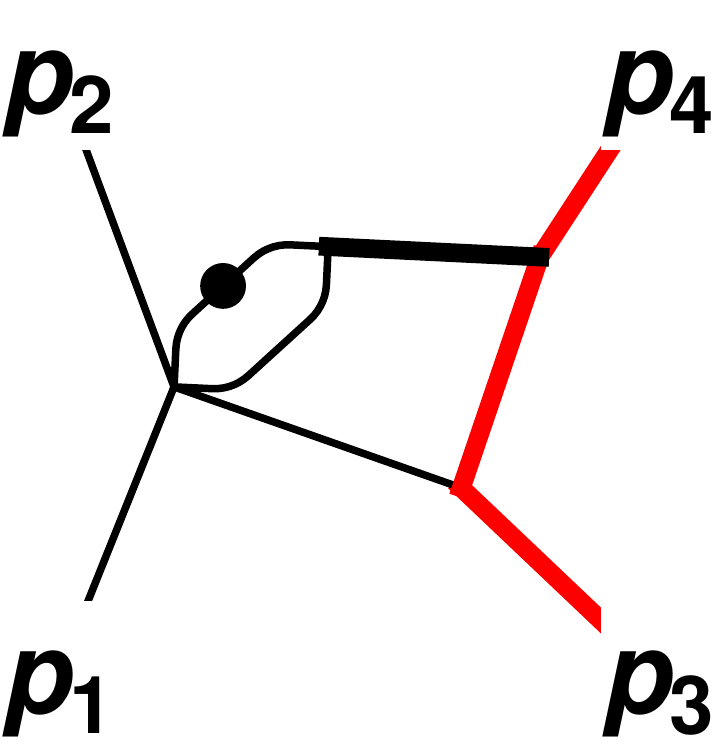}
  } \,
        \subfloat[$\mathcal{T}_{43}$]{%
    \includegraphics[width=0.105\textwidth]{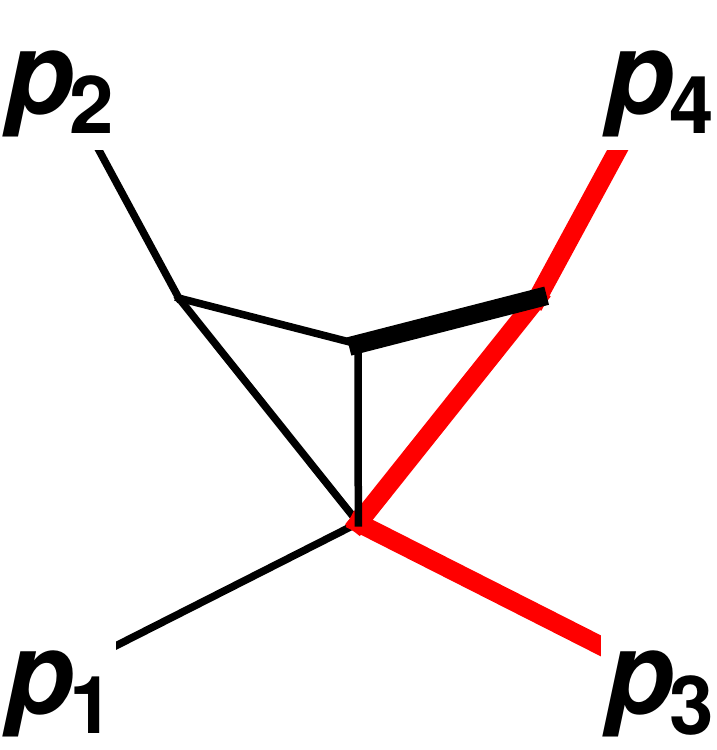}
  }\,
  \subfloat[$\mathcal{T}_{44}$]{%
    \includegraphics[width=0.105\textwidth]{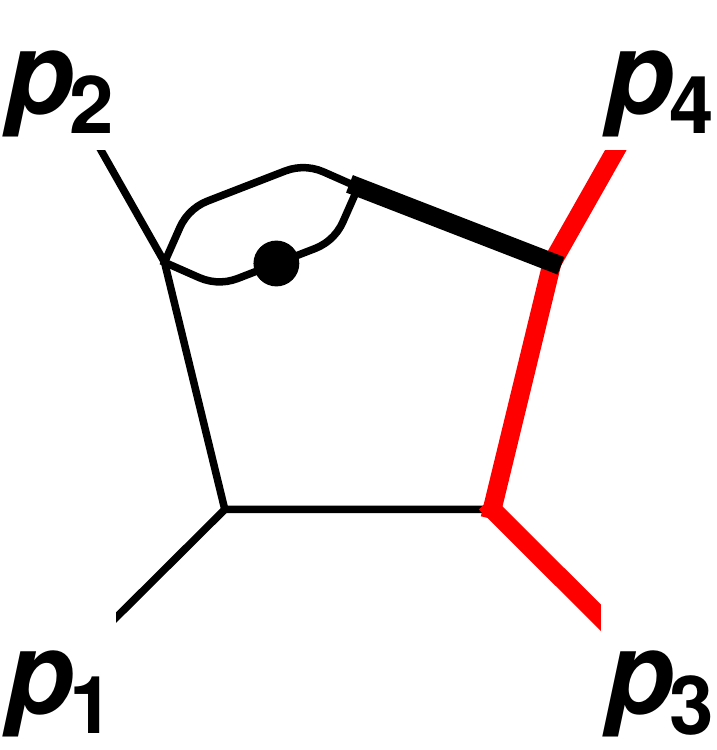}
  }\,
  \subfloat[$\mathcal{T}_{45}$]{%
    \includegraphics[width=0.105\textwidth]{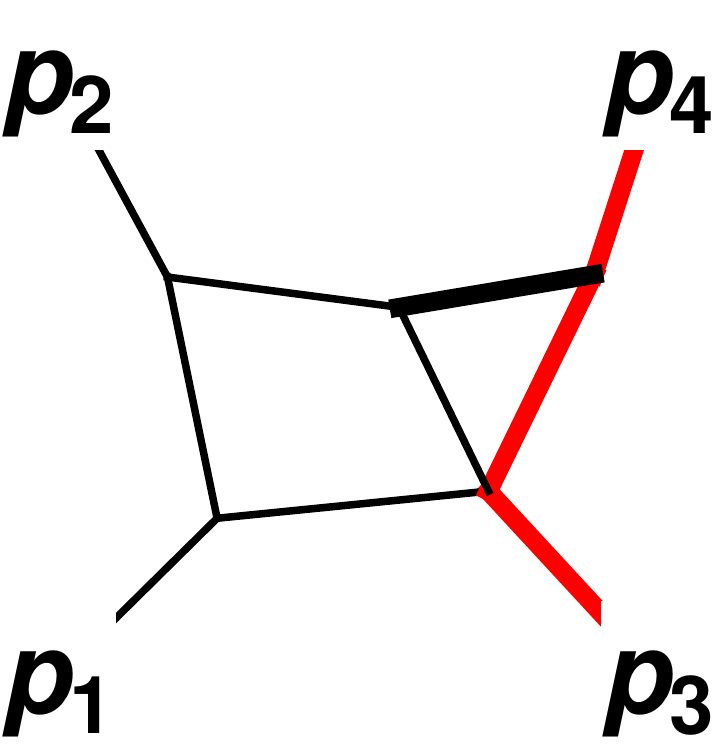}
  }\,
  \subfloat[$\mathcal{T}_{46}$]{%
    \includegraphics[width=0.105\textwidth]{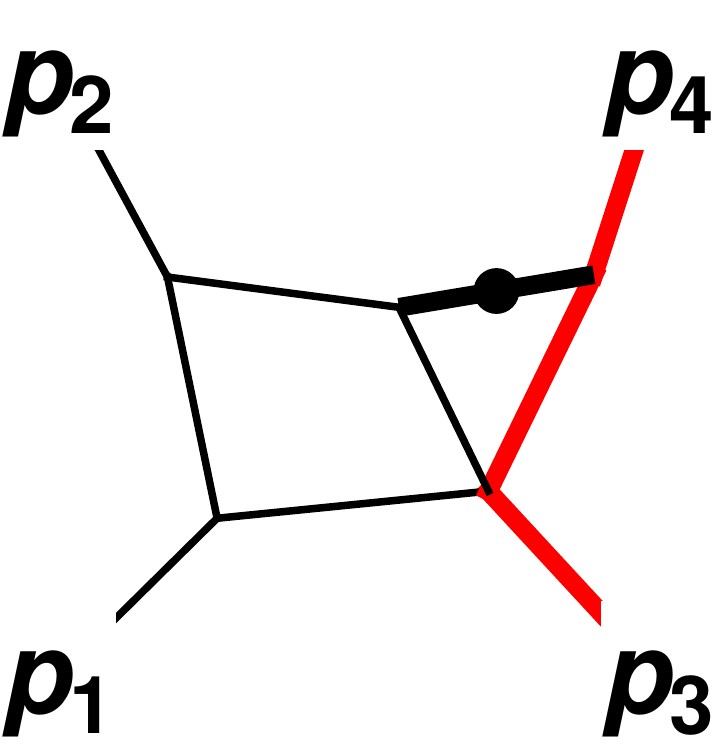}
  }\,
  \subfloat[$\mathcal{T}_{47}$]{%
    \includegraphics[width=0.105\textwidth]{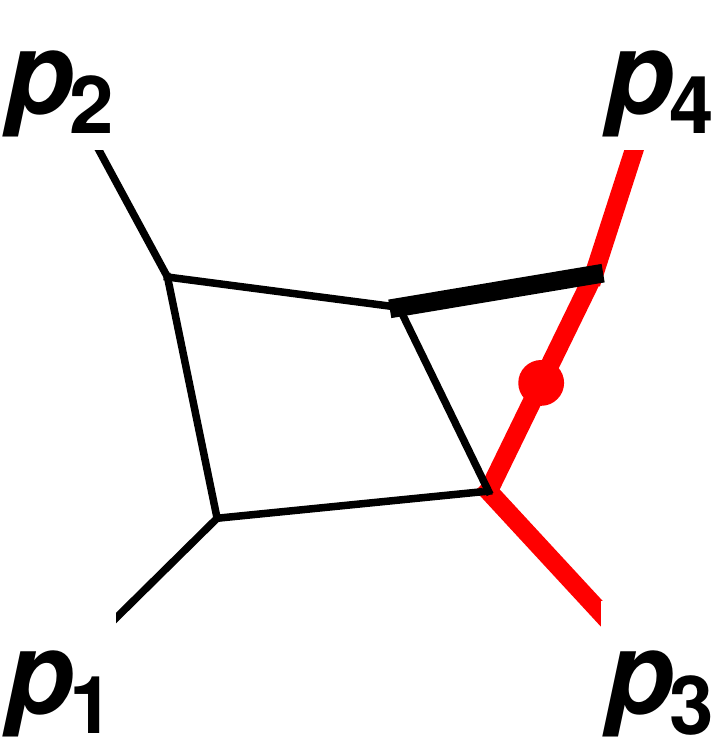}
  }\,
  \subfloat[$\mathcal{T}_{48}$]{%
    \includegraphics[width=0.105\textwidth]{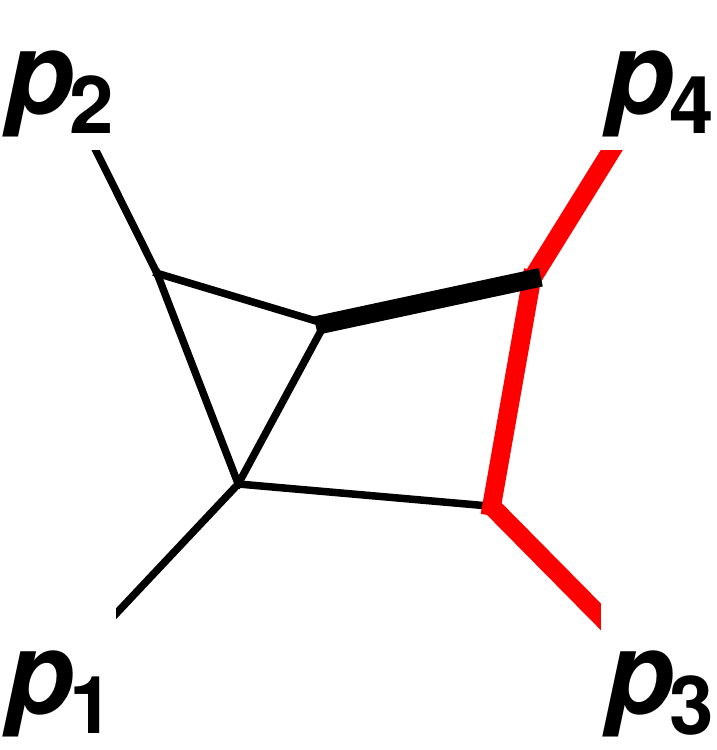}
  }\\
  \subfloat[$\mathcal{T}_{49}$]{%
    \includegraphics[width=0.105\textwidth]{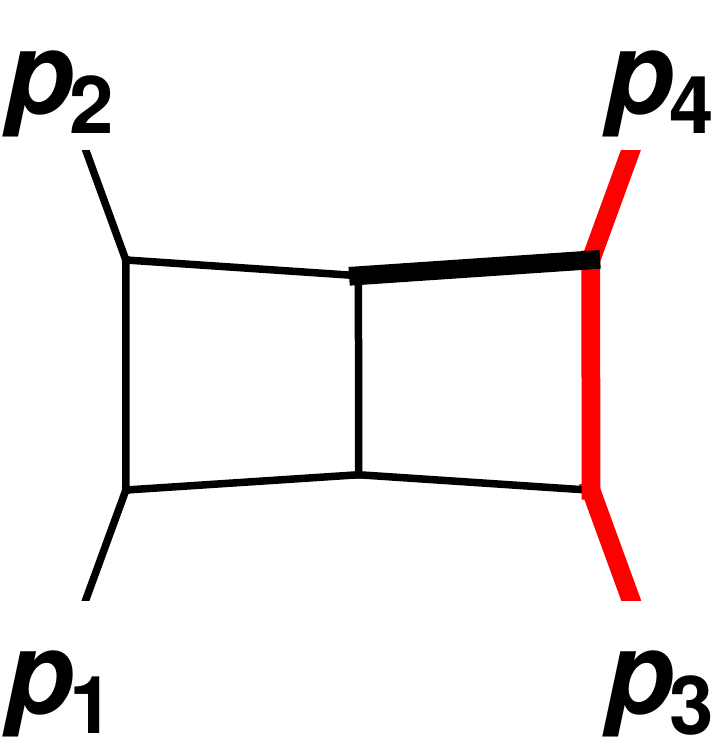}
  }\,
  \subfloat[$\mathcal{T}_{50}$]{%
    \includegraphics[width=0.105\textwidth]{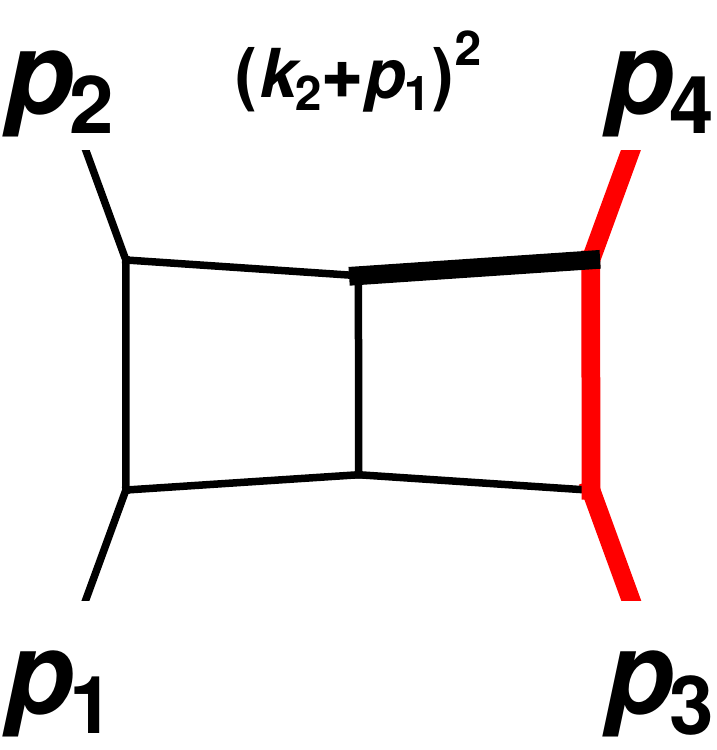}
  }\,
    \subfloat[$\mathcal{T}_{51}$]{%
    \includegraphics[width=0.105\textwidth]{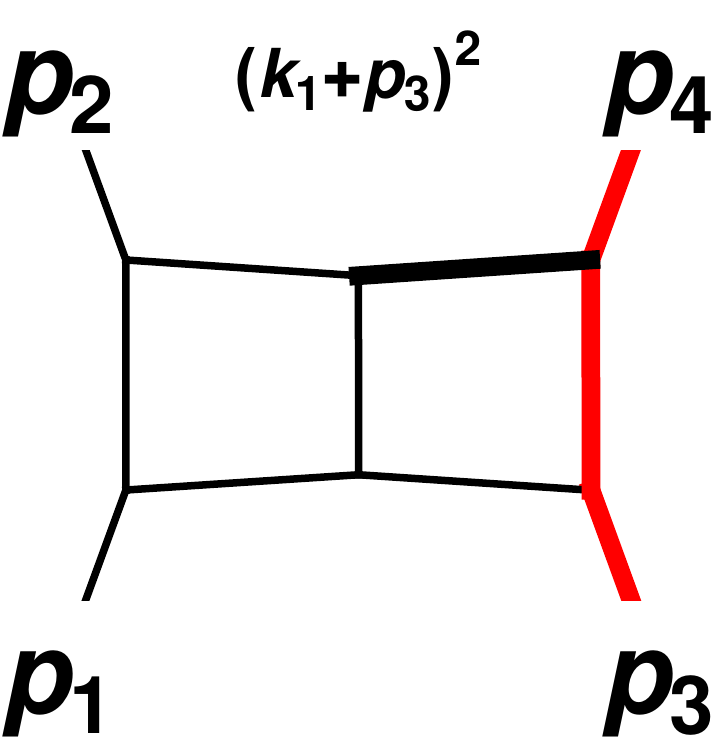}
  }
 \caption{The master integrals $\top{1\dots51}$ for the two-loop mixed QCD-QED corrections to Drell-Yan. Thin black lines represent massless propagtors, while thick black lines and red lines represent massive propagtors with mass $m_z$ or $m_l$ respectively. The dots represent additional powers of the propagator and potential numerators are written on top of each figure. 
}
 \label{fig:MIs}
\end{figure}


%% file: tex/conclusions.tex
\section{Conclusions}
The subject of this publication was the calculation of the previously unknown master integrals, needed for the two-loop mixed QCD-QED virtual corrections to the neutral current Drell-Yan process ($q\bar{q}\rightarrow l^+ l^-$). The lepton mass dependence was kept up to logarithmic terms, such that the incomplete cancellation of these potentially large contributions can be studied in cross sections which are not fully inclusive due to lepton identification cuts. This study requires the knowledge of the two-loop virtual amplitudes, whose computation is now feasible through this publication.  

The 51 master integrals were evaluated with the method of the differential equations. In particular we found a set of MIs obeying a precanonical system of differential equations, which was brought to an $\eps$-factorized form with the help of the Magnus exponential. After an expansion for small lepton masses, the boundary conditions were imposed by matching the solutions onto simpler integrals at special kinematic points, or by requiring the regularity of the solution at pseudo-thresholds. Finally the coefficients of the Taylor series around four space-time dimensions were given in terms of generalized polylogarithms up to weight four. 
